\documentclass[aps,prb,eqsecnum,showpacs, notitlepage, floats,superscriptaddress,10pt]{revtex4-1}
\usepackage{graphicx}
\usepackage{dcolumn}
\usepackage{amsmath}
\usepackage{verbatim}
\usepackage{bbold}
\usepackage{subeqnarray}
\usepackage{bm}
\usepackage{color}
\usepackage{lmodern}
\usepackage[caption=false]{subfig}
\usepackage{amssymb}
\usepackage{mathrsfs}
\usepackage{pict2e}
\usepackage{tikz}
\usepackage{mathtools}
\DeclarePairedDelimiter\bra{\langle}{\rvert}
\DeclarePairedDelimiter\ket{\lvert}{\rangle}
\DeclarePairedDelimiterX\braket[2]{\langle}{\rangle}{#1 \delimsize\vert #2}
\usepackage[breaklinks=true,colorlinks=true,linkcolor=blue,urlcolor=blue,citecolor=magenta]{hyperref}

 \newcounter{multifig}
 
 \newcommand{\multifig}{\setcounter{multifig}{\thefigure}}
 
\begin{document}
\title{Landau-Zener transitions in a qubit periodically driven in both longitudinal and transverse directions}
\author{A. B. Tchapda}
\affiliation{Mesoscopic and Multilayer Structures Laboratory, Faculty of Science, Department of Physics, University of Dschang, Cameroon}
\author{M. B. Kenmoe}
\affiliation{Mesoscopic and Multilayer Structures Laboratory, Faculty of Science, Department of Physics, University of Dschang, Cameroon}
\author{A. D. Kammogne}
\affiliation{Mesoscopic and Multilayer Structures Laboratory, Faculty of Science, Department of Physics, University of Dschang, Cameroon}
\author{L. C. Fai}
\affiliation{Mesoscopic and Multilayer Structures Laboratory, Faculty of Science, Department of Physics, University of Dschang, Cameroon}
\date{\today}

\begin{abstract}
We theoretically investigate the dynamics of a spin-qubit periodically driven in both longitudinal and transverse directions by two classical fields  respectively a radio-frequency (RF) and a microwave (MW) field operating at phase difference $\phi$. The qubit is simultaneously locally subject to a linearly polarized magnetic field which changes its sign at a degeneracy point in the longitudinal direction and remains constant in the transverse direction. We superimpose the RF and MW signals respectively to the longitudinal and transverse components of the magnetic field. The proposed model  may be used to optimize the control of a qubit in quantum devices. The various fields applied are relevant to {\it nearly-decouple} the spin-qubit from its environment, minimize decoherence effects and improve on the coherence time. The study is carried out in the Schr\"odinger and Bloch pictures. We consider the limits of weak and strong longitudinal drives set up by comparing the characteristic time of non-adiabatic transitions with the coherence time of the longitudinal drive. Expressions for populations are compared with numerics and remarkable agreements are observed as both solutions are barely discernible. 
\end{abstract}
\maketitle
\section{Introduction}\label{Sec1}
Among the various mechanisms for controlling spin flips dynamics in quantum or semi-quantum devices,  the Landau-Zener-St\"uckelberg-Majorana (LZSM)\cite{lan,zen, stu, Majorana} mechanism remains one of the few concepts that allows nearly-optimal control of population inversion at singlet-triplet anti-crossing in real/artificial devices including two- three- and multi-level systems (electrons, photons, atoms, molecules)\cite{Bulka, Cao, Ribeiro, Petta, Shev, Oliver}. This is most likely not only due to the desirable minimal number of two controllable parameters involved (the sweep velocity $v$ and the tunnel matrix element $\Delta$), the simplicity of its large positive time asymptotic solution but mainly to its minor sensitivity to certain types of noise (such as diagonal quantum or classical noises\cite{Saito2007}). It has opened a promising avenue for  implementing logical gates with high fidelity\cite{Cao} and developing quantum technologies. The LZSM mechanism  permits to realize coherent superposition of singlet and triplet spin-qubit (unit of binary quantum information)  states in double quantum dots\cite{Ribeiro}, to encode and coherently control a qubit  in the spin of a two-electron state system\cite{Ribeiro, Petta},  to design single qubit  operations for controlling superconducting qubits\cite{Shev, Oliver}, to estimate the energy gap in nanomagnets\cite{Wernsd}, to measure the decoherence time in quantum information processing\cite{Oliver, Sillanp} etc.

Although, the incommensurable list of successes ascribed to the LZSM model, several drastic drawbacks unfortunately go along with the model.  The linear drive achieves infinitely large values as the time goes to infinity. The inter-level distance between level position always remains constant and never turns off. In addition, the  model allows only a single passage at an avoided energy-level and is not in general desirable experimentally to infer information about the complex dynamics of  a system bathing in its environment. Thus, it is commonly required that the system traverses  several crossings or the same crossing several times back and forth. One way of achieving these, consists of periodically changing one of the control parameters of the system (detuning and/or Rabi frequency for instance). As a consequence, the wave function splits and evolves through different paths accumulating a phase difference. If it is driven back and passes through the same avoided level crossing, it recombines and leads to interference patterns referred to as LZSM interferences that are inspected in spectroscopy analysis to  capture the features of the system. They are also used to control the final state probability in solid states quantum devices\cite{Petta2012}. As an example, a chirped microwave has recently enabled to generate avoided level crossings in 3D transmons where these were not expected and LZSM interferences were reported\cite{sun2016}. 

The LZSM model is amended in various ways. Remarkably, changing the detuning as $vt\to A\cos(\omega t)$ (where $A$ and $\omega$ are respectively the amplitude and frequency of the longitudinal drive) is the archetype of several passed and ongoing discussions carried out both theoretically and/or experimentally\cite{Petta2012, Vitanov1997, Kayanuma2000, Wubs, Child, Kayanuma2001}. Thus, hallmark quantum effects such as cascaded LZSM transitions\cite{Oliver} (that are indicators revealing that the periodically driven system intrinsically goes through several crossings), multiphoton transitions, coherent destruction of tunneling\cite{Grossmann}  and interference patterns (indicating that passing through crossings, the system splits and recombines several times) are observed. This renormalization receipt applied to the LZSM model has stimulated intensive theoretical works. Among other things, by superimposing the detuning as $vt\to vt+A\cos(\omega t)$ and preserving the gap, cascaded LZSM transitions are observed in both the high- and low- frequency regimes of the periodic modulation for weak couplings\cite{Kayanuma2000}. These reveal that the electric field is quasiquantized in the high frequency regime while the low-frequency regime leads to real crossings. The LZSM mechanism by periodic drive is therefore a useful tool for creating energy-level crossings in systems where these should not have been necessarily found.  In the same vein, by preserving the detuning while renormalizing the energy gap as $\Delta\to \mathcal{A}_{f}\cos(\omega_{f} t+\phi)$ (where $\mathcal{A}_{f}$ and $\omega_{f}$ are respectively the amplitude and frequency of the transverse drive), it was shown in Refs.[\onlinecite{Wubs, Sarr2017, Mullen}] that the LZSM serves as an excellent tool for electromagnetic control of qubit. Similarly, the detuning and the gap were periodically modified in [\onlinecite{Child}] for control of qubit in nitrogen vacancy center. 

In this paper, the LZSM model is yet amended. Here, instead of renormalizing the detuning and/or the Rabi frequency, as in previous works, we superimpose two periodic drives respectively to the diagonal and off-diagonal components of the  traditional LZSM Hamiltonian. The new scenario permits to investigate the dynamics of a two-level system subject to a magnetic field   (whose longitudinal component changes its sign at a resonance point while the transverse part remains constant) and is  simultaneously periodically driven in both longitudinal and transverse directions.  Special emphasis is put on spin qubit for potential applications in quantum information processing. This protocol enables optimizing spin-qubit control during LZSM transitions. The various fields applied may be tuned to nearly-decouple the qubit from its environment in semiconductor quantum dots, minimize decoherence and increase the coherence time of the qubit. The remaining part of the paper is organized as follows: The generic model of the study is presented in Section \ref{Sec2}. It is investigated in the Schr\"odinger picture in Section \ref{Sec3} while Section \ref{Sec4} does similar investigations in the Bloch picture. In Section \ref{Sec5}, we compare our results with previous ones and conclude the paper in Section \ref{Sec6} by highlighting our main achievements.

\section{Theoretical Model}\label{Sec2}

\begin{figure*}[t]
	\vspace{-0.5cm}
	\multifig
	\begin{center} 
		\hspace{-1cm}
		\begin{tabular}{ccc}
			\includegraphics[width=10cm, height=7cm]{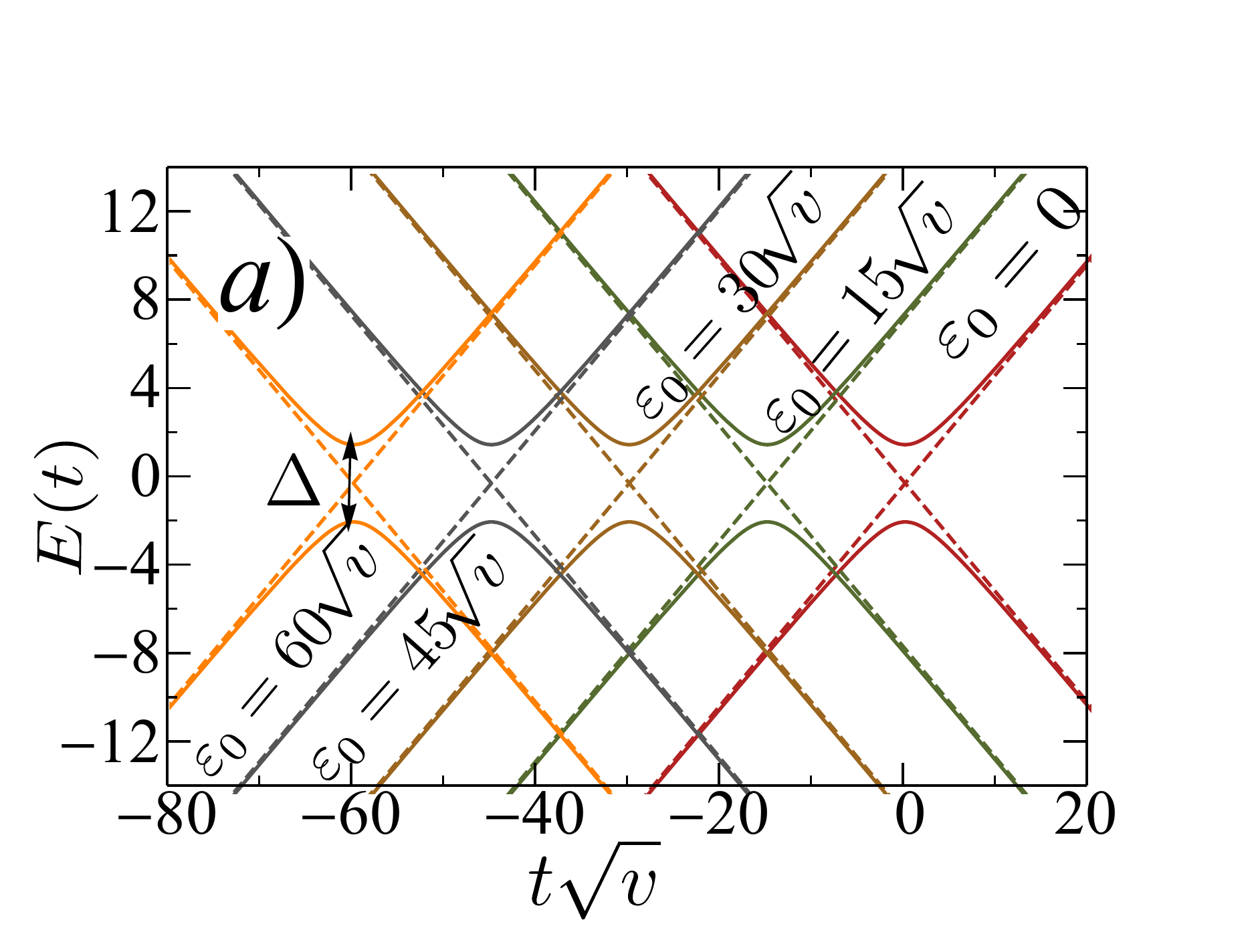}\hspace{-1.2cm}
		\end{tabular}
		\hspace{-0.5cm}
		\begin{tabular}{ccc}
			\includegraphics[width=4cm, height=4.5cm]{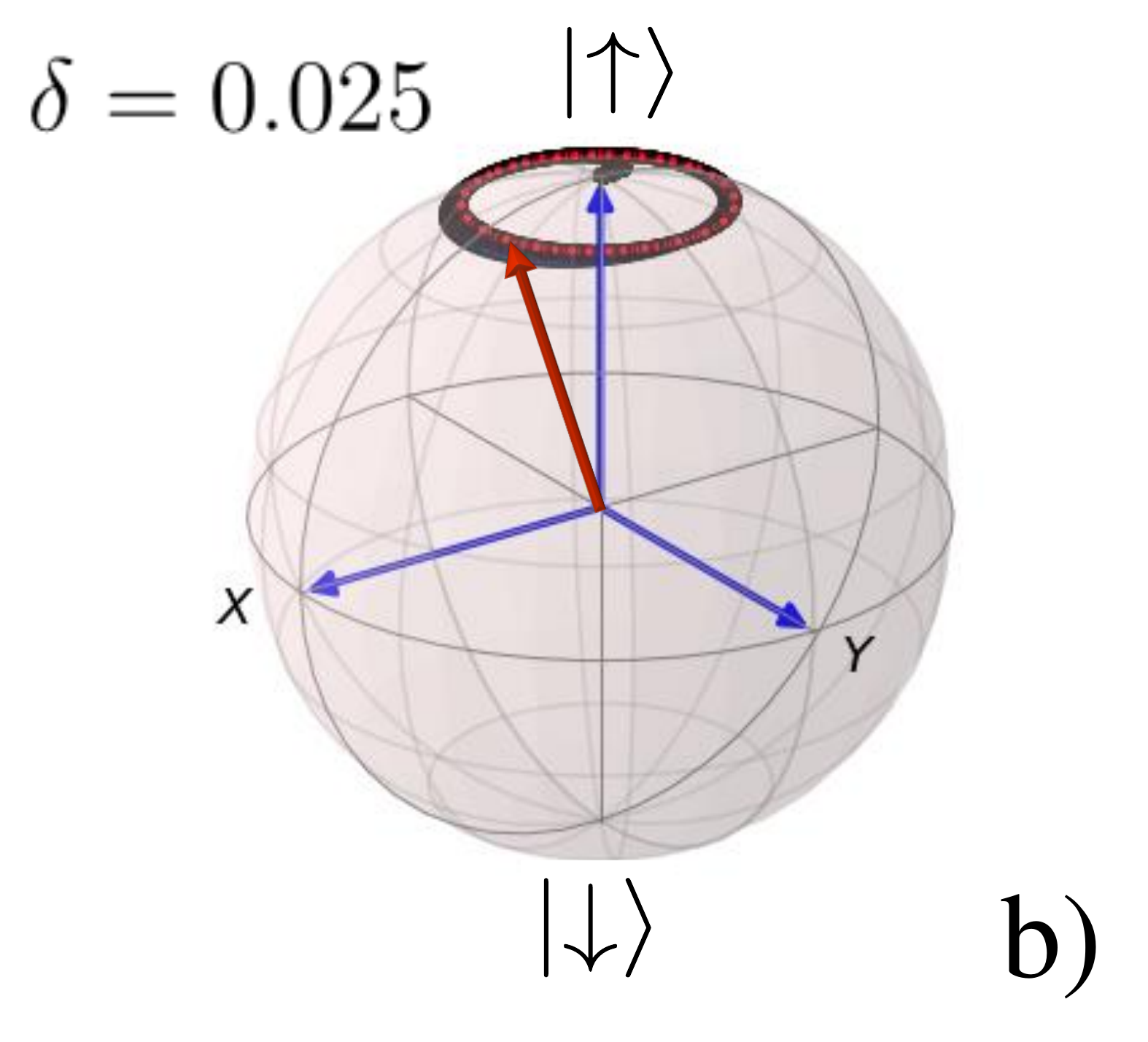}\\\vspace{-0.5cm}
			\includegraphics[width=4cm, height=4.5cm]{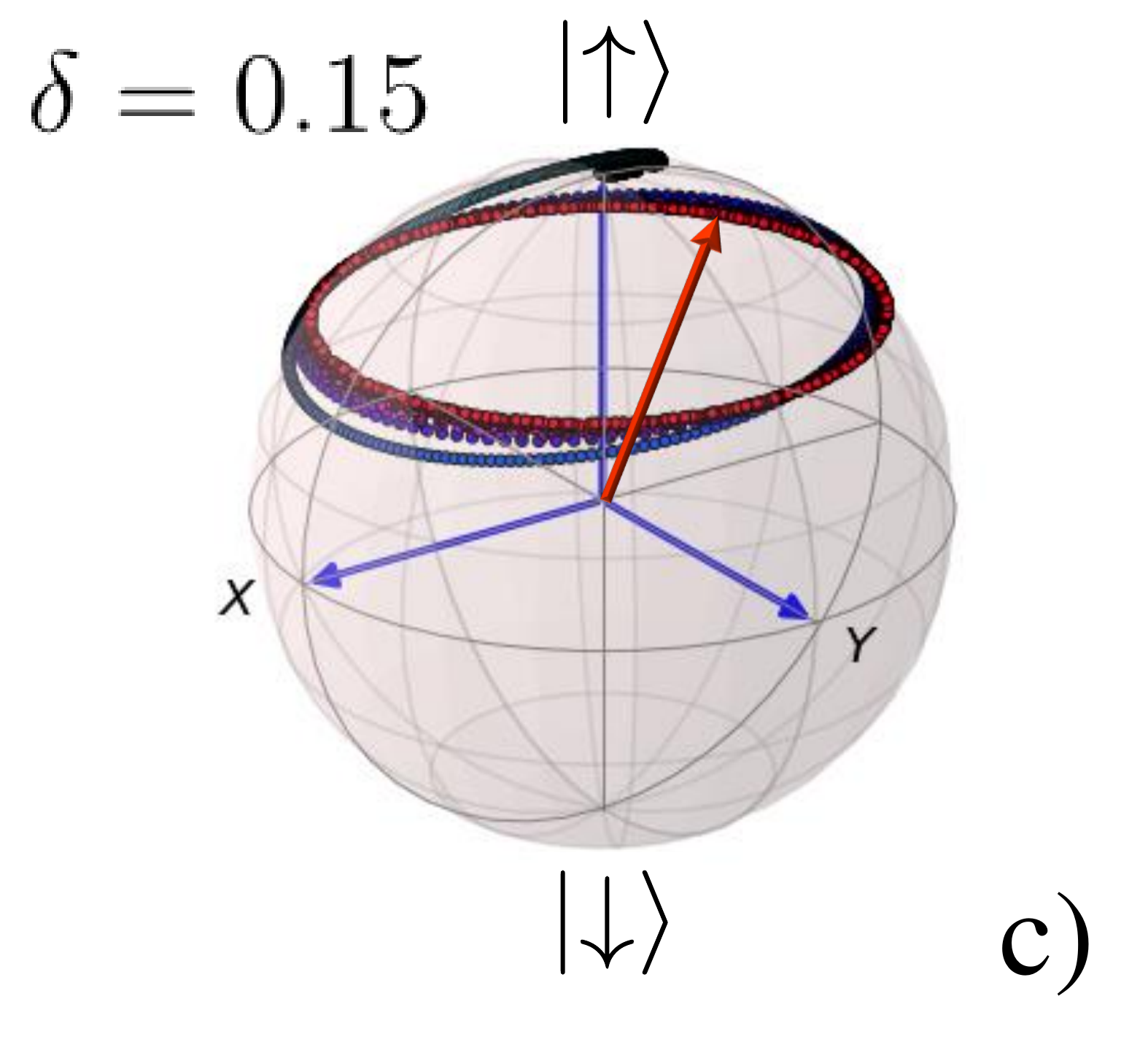}
			\vspace{-0.0cm}
		\end{tabular}
		\hspace{-1.cm}
		\begin{tabular}{ccc}
			\includegraphics[width=4cm, height=4.5cm]{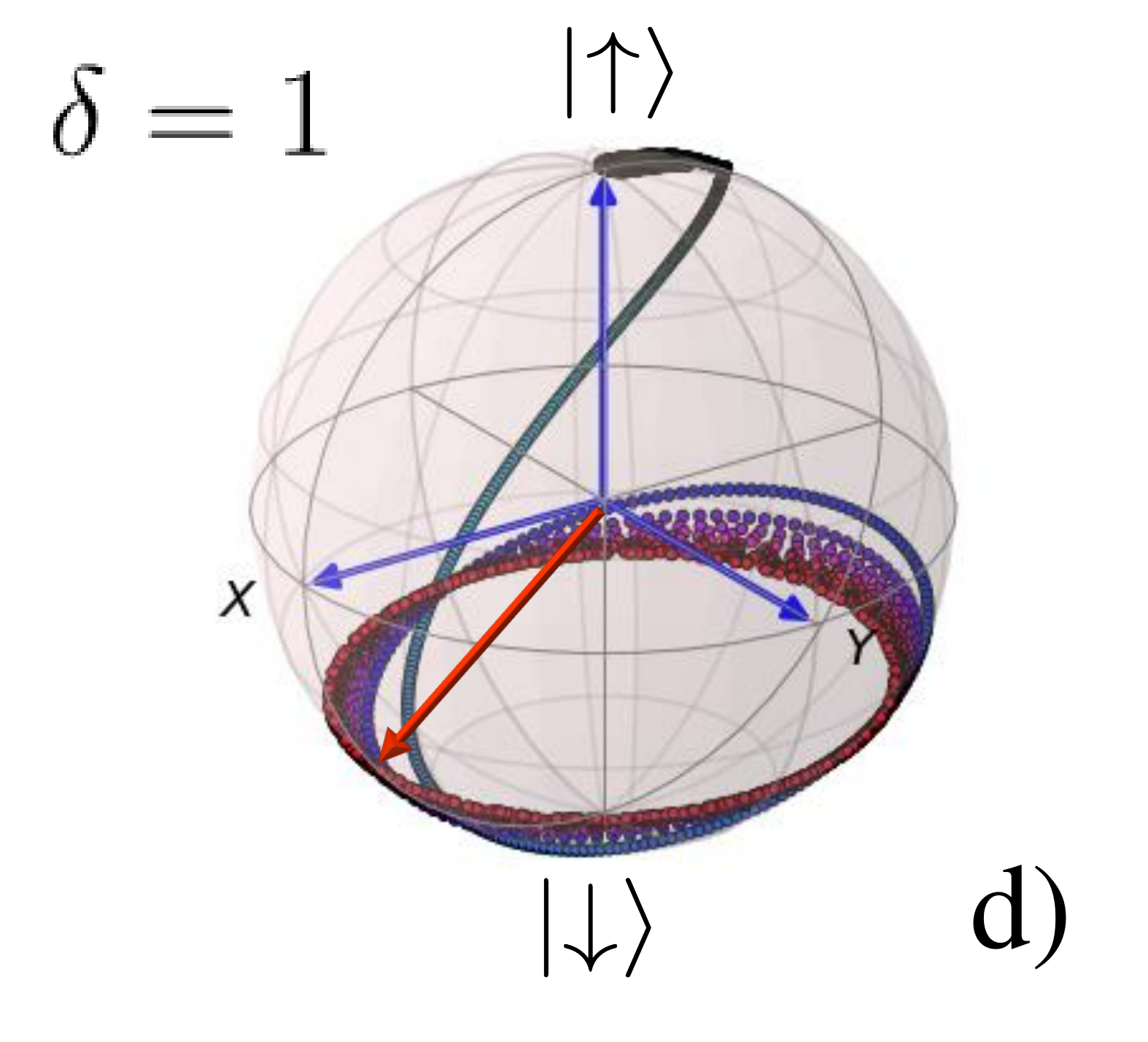}\\\vspace{-0.5cm}
			\includegraphics[width=4cm, height=4.5cm]{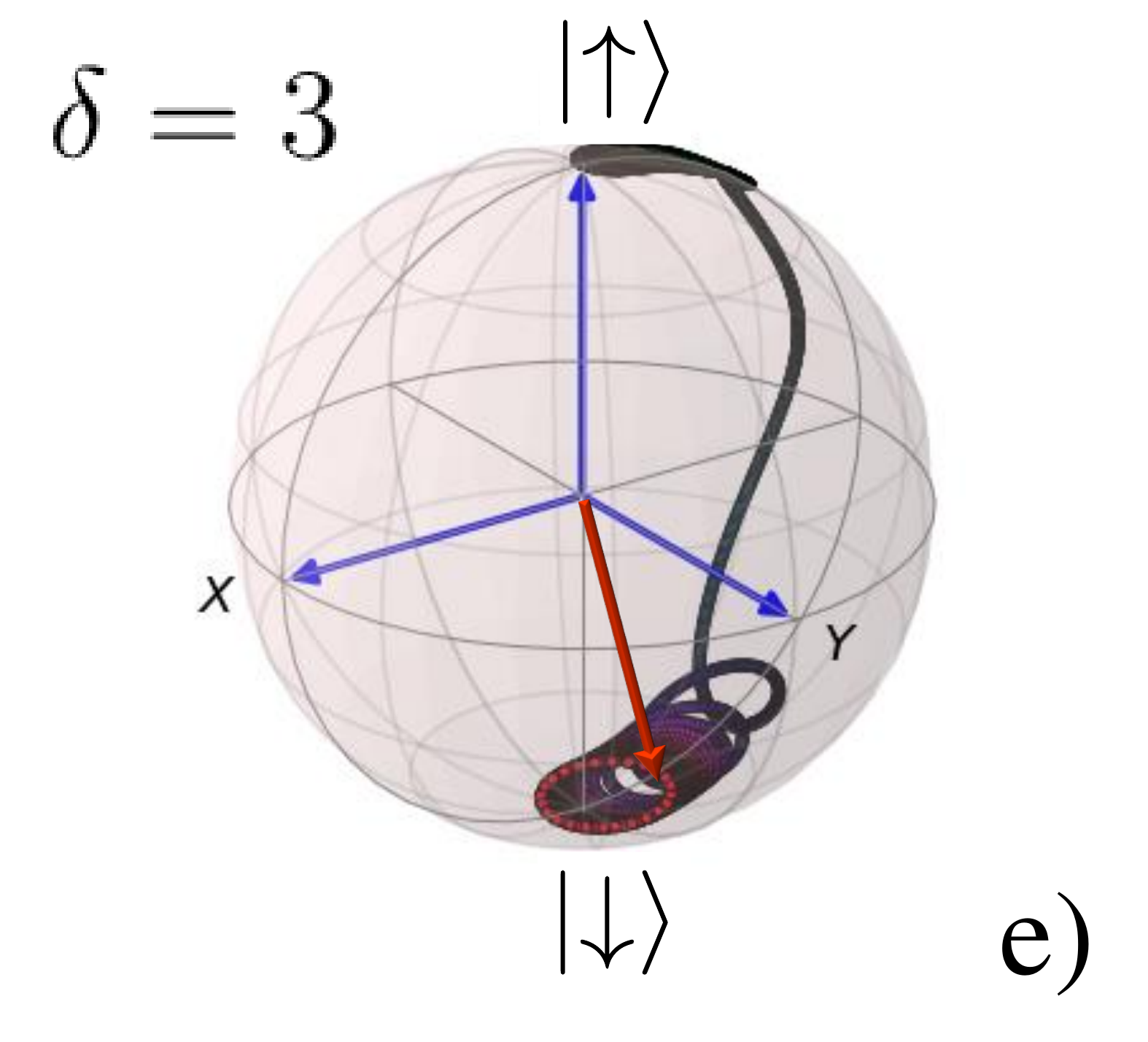}
			\vspace{-0.0cm}
		\end{tabular}
	\end{center}
	\vspace{-0.2cm}
	\caption{ a). Energy diagram for the LZSM model (\ref{equ2}) plotted for various values of the static part $\varepsilon_{0}$ in the detuning. We have considered for calculations $\delta=\Delta^2/v=3.5$. 
		We clearly see from here that by continuously changing $\varepsilon_{0}>0$, this shifts the avoided level crossing to the left of $t=0$. The shift occurs at the right of $t=0$ for $\varepsilon_{0}<0$. b) and c) on one hand d) and e) on the other hand respectively describe non-adiabatic and adiabatic LZSM  transitions in the Bloch sphere. During non-adiabatic transitions, the Bloch vector (red arrow) leaves the north pole at time $t_0=-\infty$ but remains near the equatorial plane in the south pole. This indicates that the system in average stays in its initial state. During adiabatic transitions, the Bloch vector leaves the north pole and migrate to the south pole where it dwells. This testifies a spin-flip or population inversion. The time is in the unit of $1/\sqrt{v}$.}
	\label{Figure1} 
\end{figure*}

 The two levels of the qubit are linearly swept in the quantization direction by a linearly changing-in-time magnetic field. They come close and cross offering the possibility for non-adiabatic transitions between bare states (states of the system in the absence of coupling)  at the level crossing in the fast drive regime. Let us maintain the coupling between levels as constant throughout the course of time; the qubit levels hybridize at the level crossing rather creating an avoided level crossing. This offers the possibility for adiabatic transitions in the slow sweep limit.  Now, let us subject the qubit to two classical fields such that interactions between the dipole moment of the qubit and the classical radiation reads $-\hat{\mathbf{D}}\cdot \mathbf{E}(t)
$ where $\hat{\mathbf{D}}=\hat{d}_{x}e_{x}+\hat{d}_{z}e_{z}$ and $\mathbf{E}(t)=E_{\rm MW}(t)e_{x}+E_{\rm RF}(t)e_{z}$ are respectively the dipole moment operator and the electric field vector; $e_{x}$ and $e_{z}$ being polarization vectors. In the dipole moment, and rotative-wave approximations the minimal model which describes this setup is globally of the form ($\hbar=1$ hereafter)
\begin{eqnarray} \label{equ1} 
\mathcal{H}(t)=
\mathcal{H}_{\rm LZ}(t)+\mathcal{H}_{\rm RF}(t)+\mathcal{H}_{\rm MW}(t).
\end{eqnarray}
The first term describes the spin vector $\vec{S}$ of the qubit coupled to the linearly varying-in-time magnetic field $\vec{b}_{\rm LZ}(t)=[\Delta, 0, vt]^{T}$ (the traditionally known LZSM effect) where $v>0$ is the constant sweep velocity of the control protocol and $\Delta$ the strength of coupling between the bare states ($T$ designates hereafter the transposed vector). This interaction writes $\vec{S}\cdot\vec{b}_{\rm LZ}(t)$ or
\begin{eqnarray} \label{equ2} 
\mathcal{H}_{\rm LZ}(t)=
\frac{\varepsilon(t)}{2}\boldsymbol{\mathrm{\sigma}}_{z}+\frac{\Delta}{2}\boldsymbol{\mathrm{\sigma}}_{x},
\end{eqnarray}
where the detuning is linearized at the vicinity of $t=0$ as $\varepsilon(t)=\varepsilon_{0}+(d\varepsilon(t)/dt|_{t=0})t$ and $d\varepsilon(t)/dt|_{t=0}\equiv v$. Here, $\boldsymbol{\mathrm{\sigma}}_{x,z}$ are pseudo-spin operators Pauli's matrices generating the two-dimensional rank 1 $su(2)$ Lie algebra $[\boldsymbol{\mathrm{\sigma}}_{\alpha},\boldsymbol{\mathrm{\sigma}}_{\beta}]=2i\epsilon_{\alpha\beta\gamma}\boldsymbol{\mathrm{\sigma}}_{\gamma}$ where $\epsilon_{\alpha\beta\gamma}$ (Levi-Civita symbols) are structure constants on the group $SU(2)$. An energy diagram associated with the model (\ref{equ1}) is presented in Fig.\ref{Figure1} for various values of the static shift $\varepsilon_{0}$. The RF and MW Hamiltonians are respectively given by
\begin{eqnarray} \label{equ3} 
\mathcal{H}_{\rm RF}(t)=\frac{A\cos (\omega t)}{2}\boldsymbol{\mathrm{\sigma}}_{z} ,
\quad
\mathcal{H}_{\rm MW}(t)=
\frac{\mathcal{A}_{f}\cos (\omega _{f} t+\phi)}{2}\boldsymbol{\mathrm{\sigma}}_{x}.
\end{eqnarray}
Here, $A$ and $\omega$ are respectively  the amplitude and frequency of the longitudinal drive. Similarly, $\mathcal{A}_{f}$ and $\omega_{f}$ are those of the transverse drive. Remark, $\mathcal{H}_{\rm  MW}(t)$ adds periodic variations onto the tunnel amplitude $\Delta$ allowing control of LZSM transitions in the spin-qubit in the transverse direction. $\varepsilon_0$ and $\Delta$ are interpreted as the principal signal (zero frequency-signal) of two bichromatic longitudinal and transverse signals $\varepsilon_{\rm RF}(t)=\varepsilon_0+A\cos (\omega t)$ and $\varepsilon_{\rm MW}(t)=\Delta+A\cos (\omega_{f}t+\phi)$ respectively. The qubit may now be regarded as a two-level system undergoing LZSM transitions and controlled by two trains of bichromatic signals coming from longitudinal and transverse directions. This is a dressed qubit; it keeps information much longer than the standard qubit\cite{Oliver}. 
For now, and for further purposes, let us rewrite the Hamiltonian as a trajectory in the basis of Pauli's matrices through the magnetic field vector $\vec{b}(t)=[\varepsilon_{\rm MW}(t), 0, vt+\varepsilon_{\rm RF}(t)]^T$ as
\begin{eqnarray} \label{equ3a} 
\mathcal{H}(t)=\frac{\vec{\boldsymbol{\mathrm{\sigma}}}\cdot \vec{b}(t)}{2},
\end{eqnarray}
where  $\vec{\boldsymbol{\mathrm{\sigma}}}=[\boldsymbol{\mathrm{\sigma}}_x,\boldsymbol{\mathrm{\sigma}}_y,\boldsymbol{\mathrm{\sigma}}_z]$.  The Hamiltonian in this form is equivalent to a classical Hamiltonian describing a gyromagnet precessing in the magnetic field $\vec{b}(t)$ or a spin vector precession in a Bloch's sphere (see Figs.\ref{Figure1} b) and \ref{Figure1} c)).  Thus, the eigenenergies of $\mathcal{H}(t)$ write
$E_{\uparrow,\downarrow}(t)=\pm b_{x}(t){\rm \csc2\varphi}(t)/2$ where $\varphi(t)=\arctan(-b_{x}(t)/b_{z}(t))/2$. Two equivalent and complementary pictures are investigated: the Schr\"odinger and Bloch's pictures.

\section{Schr\"odinger picture}\label{Sec3}

We wish to evaluate the zero-transferred population $P_{\uparrow\to\uparrow}(t)$ (survival probability) and the population transferred $P_{\uparrow\to\downarrow}(t)$ (transition probability). Quantum mechanics tells us that our goal compulsory passes  through the time-dependent Schr\"odinger equation (TDSE) (or its equivalent Bloch's form)
\begin{eqnarray}\label{equ9} 
i\frac{d}{dt}|\Phi (t)\rangle=\mathcal{H}(t)|\Phi (t)\rangle,
\end{eqnarray}
($\hbar=1$) subject to the initial condition $|\Phi (-\infty)\rangle=|s'\rangle$ and the constraint $\bra{\Phi (t)}\Phi (t)\rangle=1$. Let us denote as $\{\ket{\uparrow}, \ket{\downarrow}\}$ the set of eigenstates of $\boldsymbol{\mathrm{\sigma}}_{z}$. They are orthogonal ($\langle s|s'\rangle=\delta_{s,s'}$ with $s'=\uparrow, \downarrow$) and satisfy the closure relation $\sum_{s}|s\rangle\langle s|=\hat{\mathbf{1}}$ (where $\hat{\mathbf{1}}$ is a $2\times2$ unit matrix). Thus, $\ket{\uparrow}=[1,0]^{T}$ and $\ket{\downarrow}=[0,1]^{T}$ respectively correspond to the north and south poles of the Bloch sphere (see Fig.\ref{Figure1}) and form a basis for a two-dimensional Hilbert space in which quantum mechanics suggests to expand the total wave function as
\begin{eqnarray}\label{equ10}
|\Phi (t)\rangle=C_{\uparrow}(t)\exp\Big[-i\Big(\frac{vt^2}{4}+\frac{A}{2\omega}\sin(\omega t)+\frac{\varepsilon_0 t}{2}\Big)\Big]\ket{\uparrow}+
C_{\downarrow}(t)\exp\Big[i\Big(\frac{vt^2}{4}+\frac{A}{2\omega}\sin(\omega t)+\frac{\varepsilon_0 t}{2}\Big)\Big]\ket{\downarrow}.
\end{eqnarray}
Here, $C_{\uparrow}(t)=|\bra{\uparrow}\Phi(t)\rangle|$ and $C_{\downarrow}(t)=|\bra{\downarrow}\Phi(t)\rangle|$ are respectively the norm/amplitude of vectors resulting from the projections of $\ket{\Phi(t)}$ onto the directions of $\ket{\uparrow}$ and $\ket{\downarrow}$. They are also probability amplitude for detecting the spin-qubit in the states $\ket{\uparrow}$ and $\ket{\downarrow}$ respectively. Their evaluation allows estimation of populations and subsequently manipulation of the spin qubit for the fabrication of quantum devices.  Thus, for an initialization of the qubit in the state $|\Phi (-\infty)\rangle=\ket{\uparrow}$, then,  $P_{\uparrow\to\uparrow}(t)=|C_{\uparrow}(t)|^{2}$ and $P_{\uparrow\to\downarrow}(t)=|C_{\downarrow}(t)|^{2}$. For these reasons, we rewrite the wave function as $|\Phi (t)\rangle=U(t)\mathbf{C}(t)$ where $\mathbf{C}(t)=[C_{\uparrow}(t),C_{\downarrow}(t)]^{T}$ is a two-component vector probability amplitude and 
$U(t)=e^{-i\vartheta(t)\boldsymbol{\mathrm{\sigma}}_{z}/2}$ with $\vartheta(t)=(\frac{vt^2}{2}+\frac{A}{\omega}\sin(\omega t)+\varepsilon_0 t)
$ is the rotation operator that permits to rotate the TDSE (\ref{equ9}) from the Schr\"odinger to Dirac/interaction picture
\begin{eqnarray}\label{equ11} 
i\frac{d\mathbf{C}(t)}{dt}=\mathcal{H}_{\rm R}(t)\mathbf{C}(t),
\end{eqnarray}
where
\begin{eqnarray}\label{equ12} 
\mathcal{H}_{\rm R}(t)=U^{\dagger}(t)\mathcal{H}(t)U(t)-
iU^{\dagger}(t)\frac{dU(t)}{dt},
\end{eqnarray}
and where the symbol $\dagger$ denotes the Hermitian conjugate. Interestingly, $\mathcal{H}_{\rm R}(t)$ does not contain fast oscillating terms. To clearly see this, let us use the rotation laws  $U^{\dagger}(t)\boldsymbol{\mathrm{\sigma}}_{x}U(t)=e^{-i\vartheta(t)}\boldsymbol{\mathrm{\sigma}}_{+}+e^{i\vartheta(t)}\boldsymbol{\mathrm{\sigma}}_{-}$ and $U^{\dagger}(t)\boldsymbol{\mathrm{\sigma}}_{z}U(t)=\boldsymbol{\mathrm{\sigma}}_{z}$ where $\boldsymbol{\mathrm{\sigma}}_{\pm}=(\boldsymbol{\mathrm{\sigma}}_{x}\pm i\boldsymbol{\mathrm{\sigma}}_{y})/2
$ [These operators, also known as ladder operators are off-diagonal traceless  two-dimensional matrices. Together with $\boldsymbol{\mathrm{\sigma}}_{z}$ they form yet another basis for the $SU(2)$ group]. These operations unavoidably lead us to $\mathcal{H}_{\rm R}(t)=\varepsilon_{\rm MW}(t)(e^{-i\vartheta(t)}\boldsymbol{\mathrm{\sigma}}_{+}+e^{i\vartheta(t)}\boldsymbol{\mathrm{\sigma}}_{-})/2$. Given that when $t$ increases, the dominant contribution to the Hamiltonian (\ref{equ1}) comes from $vt/2$, we move to the basis of the dimensionless time
$\tau=t\sqrt{v}$, and subsequently make use of the Jacobi-Anger relation\cite{Book} $e^{ix\sin y}=\sum_{n=-\infty}^{\infty}J_{n}(x)e^{iny},
$
where $J_{n}(x)$ is the Bessel function of the first kind, of order $n$ and argument $x$. These additional operations  cast $\mathcal{H}_{\rm R}(\tau)$ into the form  
$\mathcal{H}_{\rm R}(\tau)=\sum_{n=-\infty}^{\infty}\mathcal{H}_{n}(\tau)$, which is an indication that the spin-qubit repeatedly passes through an avoided level crossing.  
This interesting fact is supported by the infinite summation that arises. On the other hand, each  passage is governed  by $\mathcal{H}_{n}(\tau)=\sum_{\alpha=-,0,+}\mathscr{H}_{n}^{\alpha}(\tau)$. The presence of a summand here also indicates that during a single passage, the qubit successively traverses three resonance points each associated with one of the possible values of $\alpha$. According to these, it appears that the RF field creates $n$ crossings ($n$ paths) and the MW field creates three subcrossings at each of the $n$ crossings generated by the RF field. Thus, at the subcrossing points marked by the index $\alpha$, the qubit evolution is ruled by the auxiliary Hamiltonian 
\begin{eqnarray}\label{equ13} 
\mathscr{H}_{n}^{\alpha}(\tau)=\mathcal{J}_{n}^{\alpha}\Big(\frac{A}{\omega}\Big)\left[ {\begin{array}{*{20}c}
	0 & e^{i[\tau+\omega_{n}^{\alpha}]^{2}/2}e^{-i\Psi_{n}^{\alpha}}\\
	e^{-i[\tau+\omega_{n}^{\alpha}]^{2}/2}e^{i\Psi_{n}^{\alpha}} & 0
	\end{array} } \right],
\end{eqnarray}
 which with precision of notations, clearly describes the $n^{th}$ passage of the spin-qubit through the $\alpha^{th}$ crossing point $\tau_{n}^{\alpha}=-\omega_{n}^{\alpha}$ subject to an exponential phase jump $e^{-i\Psi_{n}^{\alpha}}$. This corresponds to $SU(2)$ LZSM transitions between two bare states coupled through a time-independent transverse signal of amplitude the effective Rabi couplings
\begin{eqnarray}\label{equ14}
\mathcal{J}_{n}^{\alpha}\Big(\frac{A}{\omega}\Big)=\frac{\Delta_{\alpha}}{4\sqrt{v}}J_{n}\Big(\frac{A}{\omega}\Big),
\end{eqnarray}
and of frequencies
\begin{eqnarray}\label{equ15}
\Psi_{n}^{\alpha}=\frac{(\omega_{n}^{\alpha})^{2}}{2}-\phi_{\alpha},
\end{eqnarray}
where 
\begin{eqnarray}\label{equ16}
\omega_{n}^{\alpha}=[\varepsilon_0+n\omega+\alpha\omega_{f}]/\sqrt{v}.
\end{eqnarray}
In the process of describing the double periodic drive in this picture through equations written in compact form, we have defined the three-valued coupling $\Delta_{\alpha}$ ($\alpha=0,+,-$) such that $\Delta_{0}=2\Delta$ and $\Delta_{+}=\Delta_{-}=\mathcal{A}_{f}$. The phase shifts are substituted as $\phi_{+}=-\phi_{-}=\phi$ and $\phi_{0}=0$.
By simply noticing that the Hamiltonian (\ref{equ13}) transforms via a phase gate $\mathbf{F}_{n}^{\alpha}=e^{i\Psi_{n}^{\alpha}\boldsymbol{\mathrm{\sigma}}_{z}/2}$ as
\begin{eqnarray}\label{equ17} 
\mathbf{F}_{n}^{\alpha}\mathscr{H}_{n}^{\alpha}(\tau)\mathbf{F}_{n}^{\alpha\dagger}
=  \mathcal{J}_{n}^{\alpha}\Big(\frac{A}{\omega}\Big)\left[ {\begin{array}{*{20}c}
	0 & e^{i[\tau+\omega_{n}^{\alpha}]^{2}/2}\\
	e^{-i[\tau+\omega_{n}^{\alpha}]^{2}/2} & 0
	\end{array} } \right] ,
\end{eqnarray}
this confirms that at the $n^{th}$ passage through the $\alpha^{th}$ sub-level crossing, the spin-qubit does not only undergo LZSM transitions, but is equally subjected to an exponential phase jump $e^{i\Psi_{n}^{\alpha}}$. Such a jump may be experimentally created by simultaneously applying two magnetic fields of equal amplitude, one of envelope $\cos\Psi_{n}^{\alpha}$ along the $x$-direction and another  of envelope $\sin\Psi_{n}^{\alpha}$ along the $y$-direction such that their contribution to the Hamiltonian reads $\cos[\Psi_{n}^{\alpha}]\boldsymbol{\mathrm{\sigma}}_{x}+\sin[\Psi_{n}^{\alpha}]\boldsymbol{\mathrm{\sigma}}_{y}$. This may yet be another means to experimentally set up the protocol described by the model (\ref{equ1}).

All the above equations are exact as no approximation has been made so far. Given that the model proposed in this piece of work cannot be solved in an exact basis, some relevant approximations are considered.

\subsection{Strong RF drive}\label{Sec2.2}
\begin{figure}[]
 	\centering
 	\begin{center} 
 		\includegraphics[width=8.3cm, height=6.6cm]{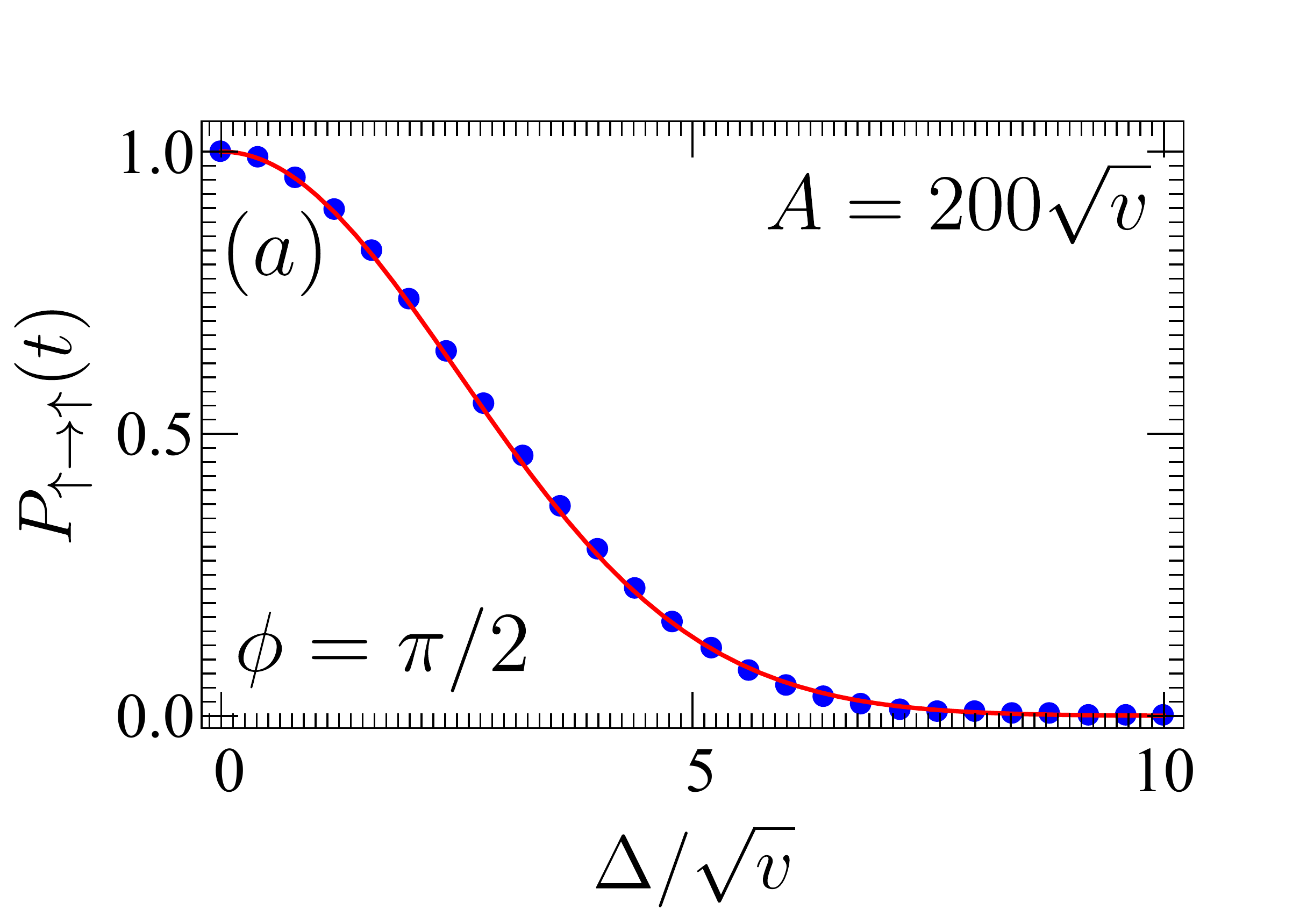}\hspace{-0.7cm}
 		\includegraphics[width=8.3cm, height=6.6cm]{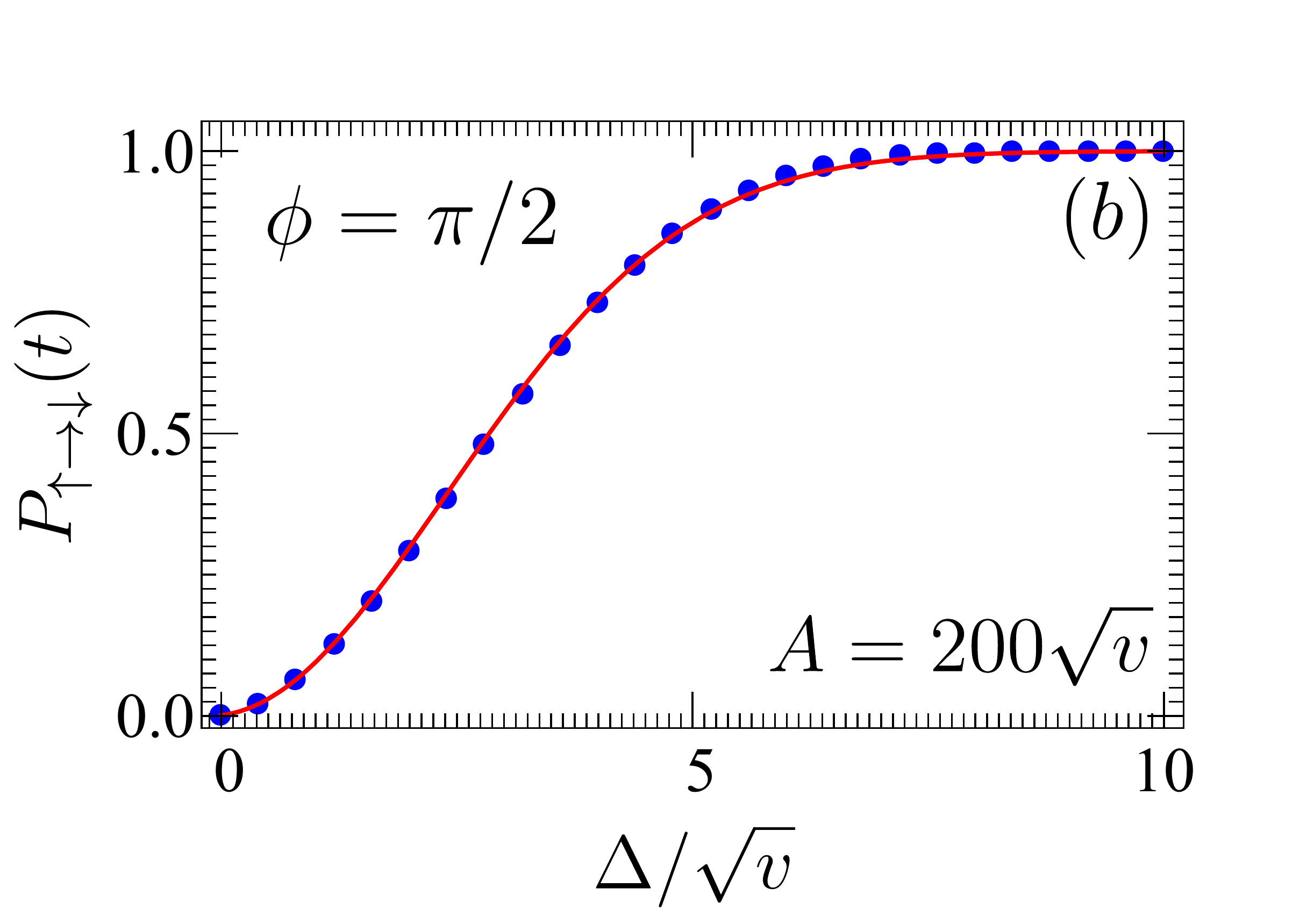}
 		\vspace{-0.5cm}
 		\caption{ Analytical (blue solid balls) versus numerical solution (red solid lines) of the Schr\"odinger equation (\ref{equ9}) in the strong RF and MW driving regimes. We have calculated the LZSM transition probability with the parameter (\ref{equ24}). We have considered $\omega=100\sqrt{v}$, $\omega_f=200\sqrt{v}$, $\mathcal{A}_f=0.08\sqrt{v}$ and $\varepsilon_0=0.0$. The integration time runs from $t\sqrt{v}=-50$ to $t\sqrt{v}=50$. The two solutions are in excellent agreement. Similar agreement is observed when the probabilities are treated as functions of $\phi$.}
 		\label{Figure7}
 	\end{center}
 \end{figure}

The ratio $A/\omega$ determines the magnitude of the coupling between diabatic states in the presence of drives. When $A/\omega=j_{n,k}$ (where $j_{n,k}$ is the $k^{th}$ zero of the Bessel function) we observe a coherent destruction of tunneling (CDT) as no population transfer occurs\cite{Grossmann}. If the RF and MW fields are tuned such that their frequencies and the amplitude of the RF field achieve very large values, the qubit stays at the resonance and performs a single LZSM transition. As a consequence, $\omega_{n}^{\alpha}=0$ and the relevant contributions to the series of Bessel functions come from 
\begin{eqnarray}\label{equ23}
n\equiv n_{\alpha}=-[\varepsilon_0+\alpha\omega_{f}]/\omega.
\end{eqnarray}
Indeed, the $\omega_{n}^{\alpha}$-dependence of the exponential phases in (\ref{equ12}) washes out. The dynamical phase accumulated vanishes in average and do not lead to interferences. The resulting Hamiltonian is nothing but the conventional LZSM with the modified LZSM parameter
\begin{eqnarray}\label{equ24}
\nonumber\delta=\sum_{\alpha,\beta}\mathcal{J}_{n_{\alpha}}^{\alpha}\Big(\frac{A}{\omega}\Big)\mathcal{J}_{n_{\beta}}^{\beta}\Big(\frac{A}{\omega}\Big)\cos[\phi_{\alpha}-\phi_{\beta}]=\Big[\mathcal{J}_{n_{0}}^{0}\Big(\frac{A}{\omega}\Big)+\sum_{\alpha\neq0}\mathcal{J}_{n_{\alpha}}^{\alpha}\Big(\frac{A}{\omega}\Big)\cos\phi_{\alpha}\Big]^2+\sum_{\alpha,\beta\neq0}\mathcal{J}_{n_{\alpha}}^{\alpha}\Big(\frac{A}{\omega}\Big)\mathcal{J}_{n_{\beta}}^{\beta}\Big(\frac{A}{\omega}\Big)\sin\phi_{\alpha}\sin\phi_{\beta}.\\
\end{eqnarray}
The time-evolution of the vector probability amplitudes  $\mathbf{m}(t)$ extracted from (\ref{equ11}) within the time interval $\tau_{i}\le \tau\le \tau_{f}$ ($\tau_{i}$ and $\tau_{f}$ are the initial and final times respectively) is ruled by the evolution operator $\hat{\mathsf{U}}(\tau_{f},\tau_{i})$ 
as\cite{Viatnov1996} 
\begin{eqnarray}\label{equ18}
\mathbf{m}(\tau_{f})=\hat{\mathsf{U}}(\tau_{f},\tau_{i})\mathbf{m}(\tau_{i}),
\end{eqnarray}
where 
\begin{eqnarray}\label{equ19}
\hat{\mathsf{U}}(\tau_{f},\tau_{i})=
\left[ {\begin{array}{*{20}c}
	a(\tau_{f},\tau_{i}) & b(\tau_{f},\tau_{i})\\
	-b^{*}(\tau_{f},\tau_{i}) & a^{*}(\tau_{f},\tau_{i})
	\end{array} } \right],
\end{eqnarray}
and where 
\begin{eqnarray}\label{equ20}
a(\tau_{f},\tau_{i})=
-\frac{i\Gamma(i\delta+1)}{\sqrt{2\pi}}\Big[D_{-i\delta}(-iz_{f})D_{-i\delta-1}(iz_{i})+D_{-i\delta}(iz_{f})D_{-i\delta-1}(-iz_{i})\Big],
\end{eqnarray}
and
\begin{eqnarray}\label{equ21}
b(\tau_{f},\tau_{i})=
\frac{\Gamma(i\delta+1)e^{-i\pi/4}}{\sqrt{2\pi\delta}}\Big[D_{-i\delta}(-iz_{f})D_{-i\delta}(iz_{i})-D_{-i\delta}(iz_{f})D_{-i\delta}(-iz_{i})\Big],
\end{eqnarray}
are Caley-Klein parameters set with the condition $|a|^{2}+|b|^{2}=1$ which ensures the unitary of $\hat{\mathsf{U}}(\tau_{f},\tau_{i})$. Here $z_{\kappa}(\tau)=\tau_{\kappa} e^{-i\pi/4}$ and $D_{\nu}(x)$ is the parabolic cylinder Weber's function of index $\nu$ and argument $x$; $\Gamma(...)$ is the Euler Gamma function\cite{Book}. Equations (\ref{equ20}) and (\ref{equ21}) are of central interest in archetypes for probing quantum systems undergoing LZSM transitions\cite{Guido}. A celebrated relation known as the LZSM formula obtained by setting the initial and final times respectively to $\tau_{i}=-\infty$ and $\tau_{f}=+\infty$ in (\ref{equ20})  is given by\cite{lan, zen, stu, Majorana}
\begin{eqnarray} \label{equ22} 
P_{\uparrow\to\uparrow}(\infty)=e^{-2\pi\delta}.
\end{eqnarray}
This formula with (\ref{equ24}) holds for arbitrary driving regime of the magnetic field. A numerical test is implemented in order to check/confirm its range of validity. We have observed that it perfectly fits the exact data of numerical calculations for large $A$, $\omega$, $\omega_f$ and arbitrary $\Delta$, $\phi$ (see Fig.\ref{Figure7}). We have however observed a discrepancy between analytical and numerical results when $\varepsilon_{0}$ achieves large values. Thus, our results hold for moderately large values of the static shift  of the detuning $\varepsilon_0$. In the extreme limit  $\varepsilon_0=0$, the LZSM parameter (\ref{equ24}) takes the form
\begin{eqnarray}\label{equ24a}
\delta=\Big[\frac{\Delta}{2\sqrt{v}}+\Big(\mathcal{J}_{Q}\Big(\frac{A}{\omega}\Big)+\mathcal{J}_{-Q}\Big(\frac{A}{\omega}\Big)\Big)\cos\phi\Big]^2+\Big(\mathcal{J}_{Q}\Big(\frac{A}{\omega}\Big)-\mathcal{J}_{-Q}\Big(\frac{A}{\omega}\Big)\Big)^2\sin^2\phi,
\end{eqnarray}
which with $Q=\omega_f/\omega$ is reminiscent to the results of the quantized form (two-mode) of (\ref{equ1})  discussed in Refs[\onlinecite{Saito2007, pok2007, Kenmoe2015}]. A correspondence between the semi-classical and quantum forms of (\ref{equ1}) may be investigated [this falls out of the scope of this paper]. 

 \subsection{Weak RF drive}\label{Sec2.3}
\begin{figure}[]
 	\centering
 	\begin{center} 
 		\includegraphics[width=8.3cm, height=6.6cm]{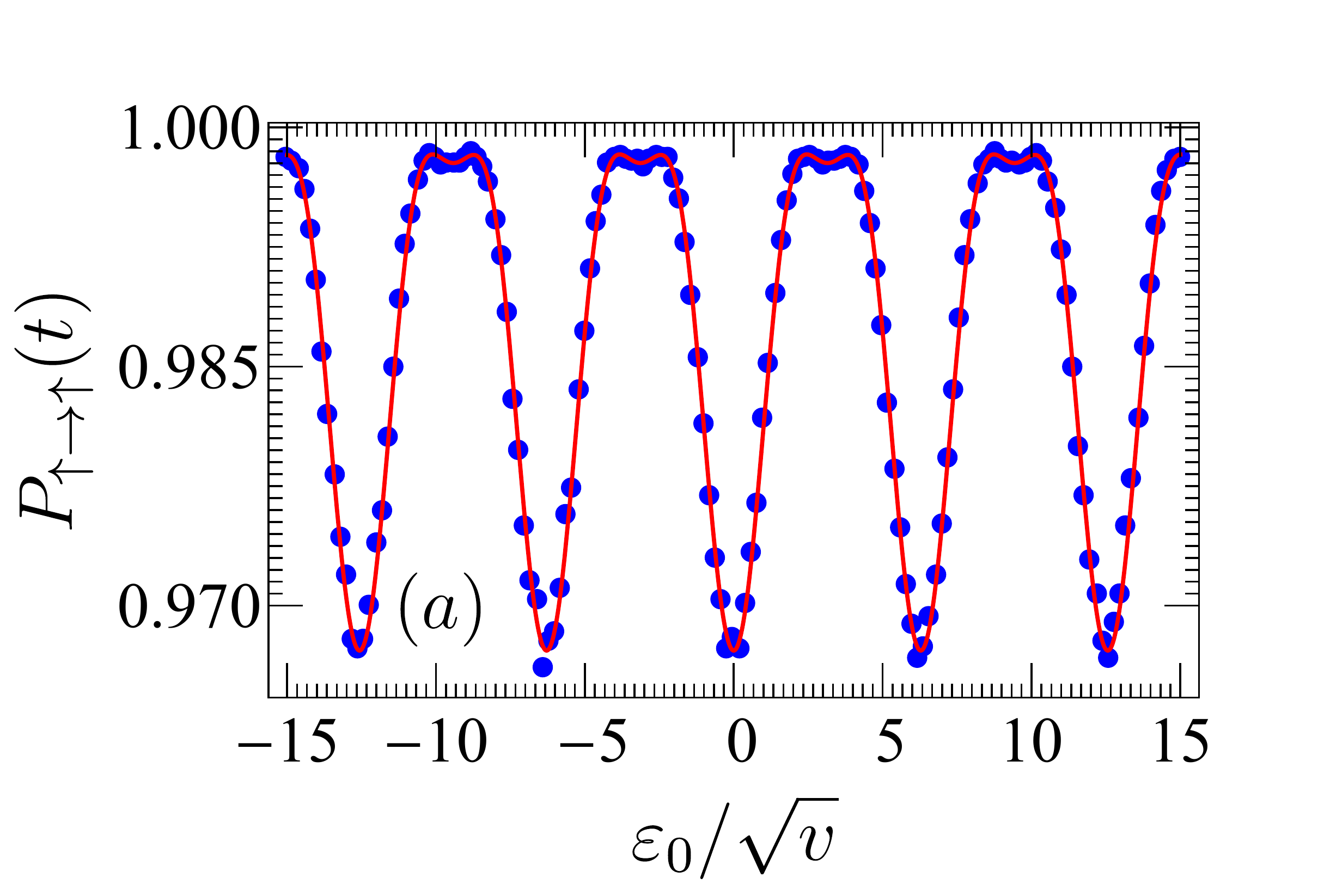}\hspace{-0.7cm}
 		\includegraphics[width=8.3cm, height=6.6cm]{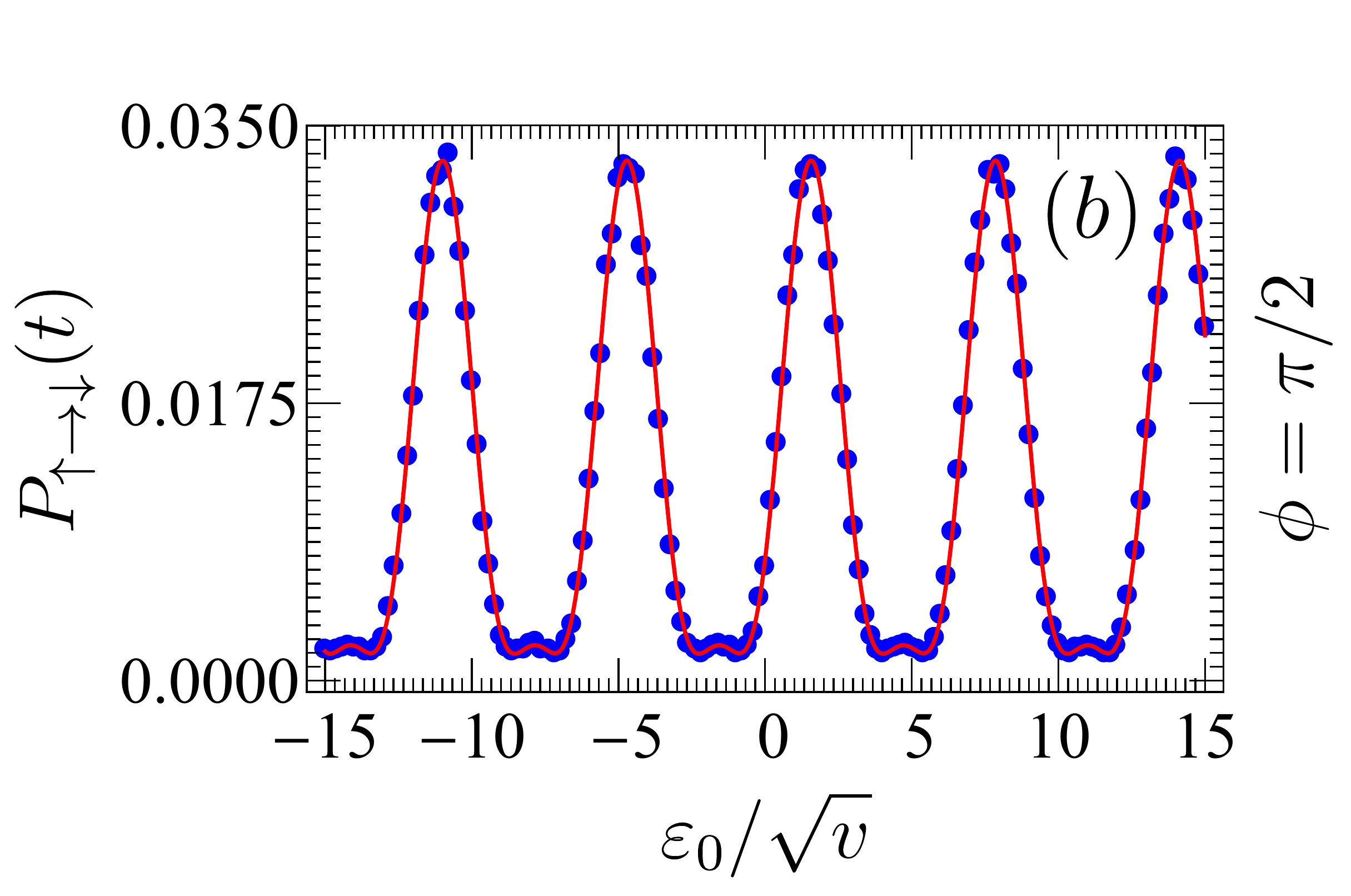}\\\vspace{-1.0cm}
 		\includegraphics[width=8.3cm, height=6.6cm]{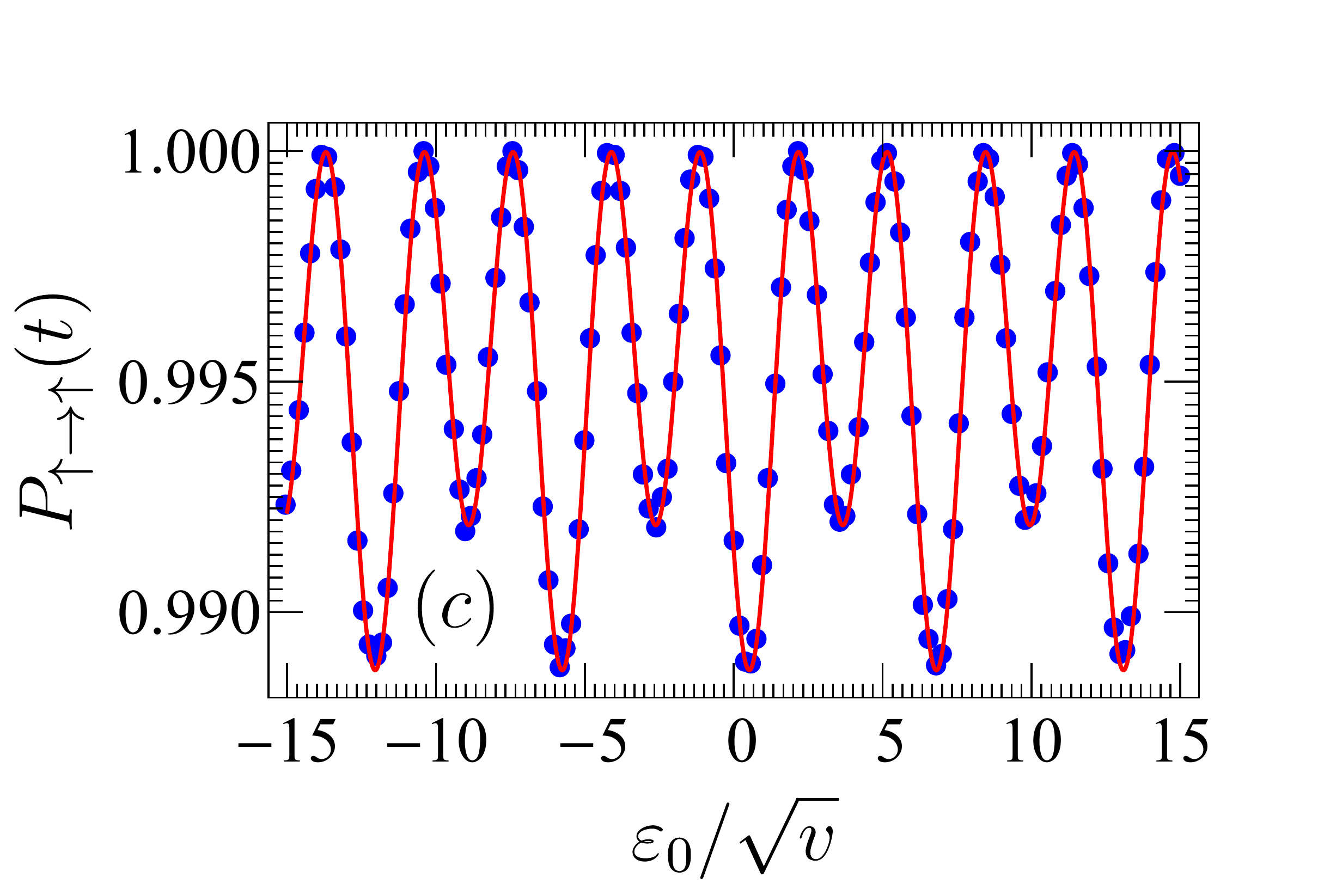}\hspace{-0.7cm}
 		\includegraphics[width=8.3cm, height=6.6cm]{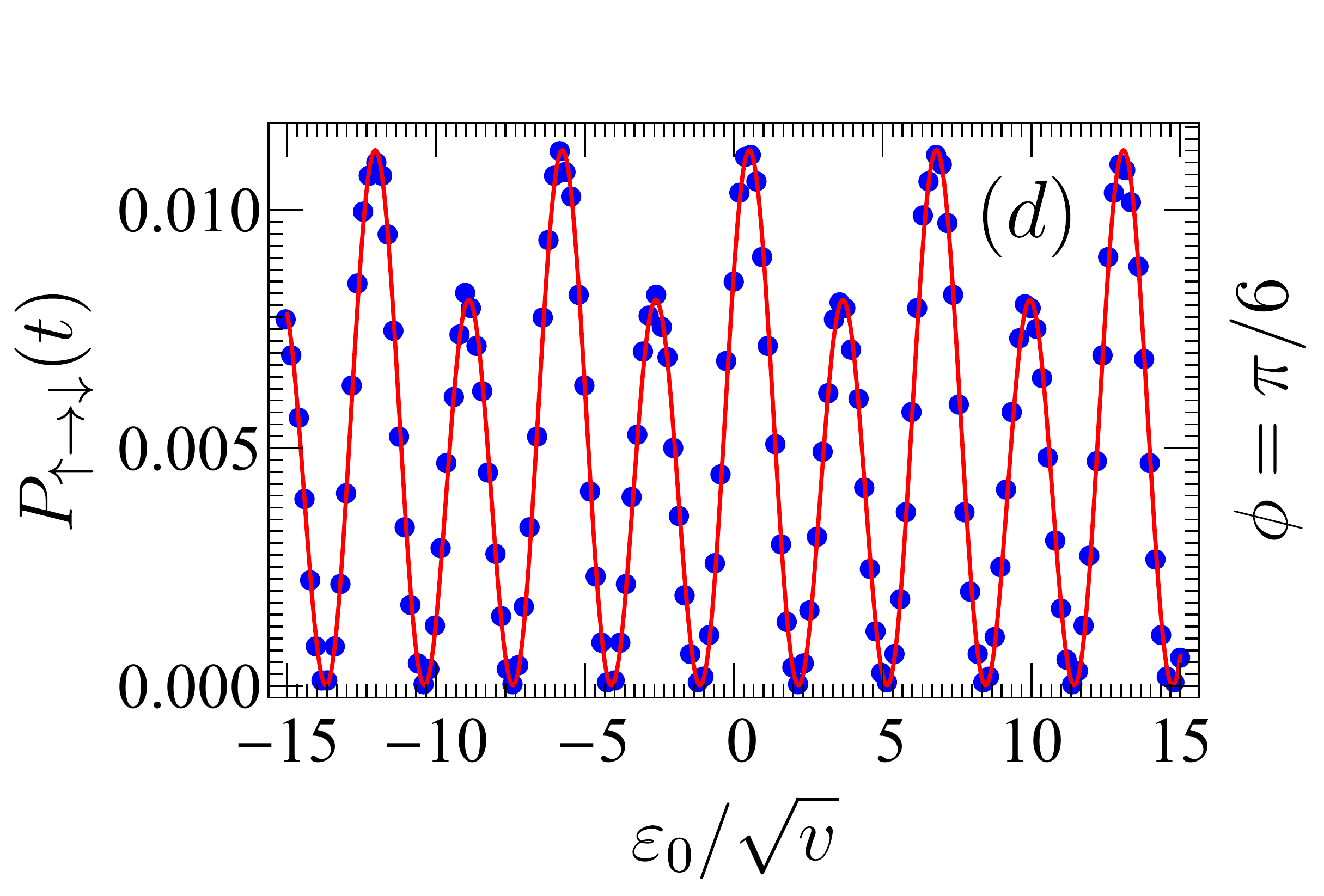} 
 		\vspace{-0.5cm}
 		\caption{ Comparison between analytical solutions (\ref{equ32a}), (\ref{equ32b}) and numerical solutions of the Schr\"odinger equation (\ref{equ9}). For calculations, $t_{0}\sqrt{v}=-50$ and $t\sqrt{v}=50$; (upper panel) $A/\sqrt{v}=1$, $\omega/\sqrt{v}=50$, $\mathcal{A}_{f}/\sqrt{v}=0.08$, $\omega_{f}/\sqrt{v}=1$ and $\Delta/\sqrt{v}=0.07$; (lower panel) $A/\sqrt{v}=29$, $\omega/\sqrt{v}=100$, $\mathcal{A}_{f}/\sqrt{v}=0.08$, $\omega_{f}/\sqrt{v}=1$ and $\Delta/\sqrt{v}=0.0075$. Blue solid balls are exact numerical results while red solid lines are analytical results. We observe a remarkable agreement between analytical and numerical results in the indicated limits ($A/\omega\ll1$ and $\Delta_{\alpha}/\sqrt{v}\ll1$) given that both results are barely discernible.}
 		\label{Figure2a}
 	\end{center}
 \end{figure}

It is worth noticing that each of the three phases in the right hand side of Eq.(\ref{equ13}) obeys a quadratic dependence and thus contribute for each $n$ only at relevant times $\tau_{n}^{\alpha}=-\omega_{n}^{\alpha}$ (crossing points) where the phases are stationary. The exact solution is approximated by considering the dominant contributions occurring at crossing times. Let $N$ be the total number of LZSM transitions executed by the qubit from $n=-\infty$ at initial time $\tau_{i}=-\infty$ to $n=+\infty$ at final time $\tau_{f}=+\infty$.
Let $\mathbf{U}(\tau_{f}, \tau_{i})$ describes its full time-evolution from $\tau_{i}$ to $\tau_{f}$ by propagating the  vector probability amplitude as $\mathbf{C}(\tau_{f})=\mathbf{U}(\tau_{f}, \tau_{i})\mathbf{C}(\tau_{i})$. As the system is continuously driven back and forth about an avoided level crossing, its full evolution can be subdivided into $N$ interconnected individual evolutions $\mathbf{U}_{\kappa}(\tau_{\kappa}, \tau_{\kappa-1})$ such that 
\begin{eqnarray}\label{equ25}
\mathbf{U}(\tau_{f}, \tau_{i})=\prod_{\kappa=N}^{1}\mathbf{U}_{\kappa}(\tau_{\kappa}, \tau_{\kappa-1}),
\end{eqnarray}
where $\tau_{f}=\tau_{N}=+\infty$, $\tau_{i}=\tau_{0}=-\infty$ and $\mathbf{C}(\tau_{0})=|s'\rangle$ with $s'=\uparrow, \downarrow$. Our goal then is to construct $\mathbf{U}(\tau_{f}, \tau_{i})$ i.e. the propagator $\mathbf{U}_{\kappa}(\tau_{\kappa}, \tau_{\kappa-1})$ of sub-evolutions. Let us consider an evolution within an arbitrary time interval $\tau_{\kappa-1}\le \tau\le \tau_{\kappa}$ with $\tau_{\kappa-1}\neq \tau_{i}$. Thus, the relevant vector probability amplitude $\mathbf{C}(\tau_{\kappa})=\mathbf{U}_{\kappa}(\tau_{\kappa}, \tau_{\kappa-1})\mathbf{C}(\tau_{\kappa-1})$. Denoting as $\mathbf{C}_{\kappa}(\tau)=\mathbf{C}(\tau_{\kappa})$ the vector probability amplitude in the time-domain  around the $\kappa^{th}$ avoided level crossing, it can be shown that $\mathbf{U}_{\kappa}(\tau_{\kappa}, \tau_{\kappa-1})\equiv\mathbf{U}_{\kappa}(\zeta)$ obeys the $\zeta$-dependent Schr\"odinger equation
\begin{eqnarray}\label{equ26}
i\frac{d\mathbf{U}_{\kappa}(\zeta)}{d\zeta}=\sum_{\alpha=-,0,+}\mathscr{H}_{\kappa}^{\alpha}(\zeta)\mathbf{U}_{\kappa}(\zeta).
\end{eqnarray}
This equation is solved under the assumption that the system starts off at time $\zeta=\tau_{\kappa-1}$ and stops at $\zeta=\tau_{\kappa}$. Let us now introduce the transfer matrix $\mathsf{S}_{\alpha}(\zeta)$  as solution to the $\zeta$-dependent Schr\"odinger equation
\begin{eqnarray}\label{equ27}
 i\frac{d\mathsf{S}_{\alpha}(\zeta)}{d\zeta}=\mathscr{H}_{\kappa}^{\alpha}(\zeta)\mathsf{S}_{\alpha}(\zeta).
 \end{eqnarray}
Assuming that the $\mathsf{S}_{\alpha}(\zeta)$-transformation weakly affects the Hamiltonian $\mathscr{H}_{\kappa}^{\alpha}(\zeta)$ in the limit $\mathcal{J}_{\kappa}^{\alpha}(A/\omega)\ll1$, then the Baker-Campbell-Haursdorff formula\cite{Cohen} for expansion of exponential operators allows us to write $\mathsf{S}_{\alpha}(\zeta)\mathscr{H}_{\kappa}^{\beta}(\zeta)\mathsf{S}_{\alpha}^{\dagger}(\zeta)\approx\mathscr{H}_{\kappa}^{\beta}(\zeta)+\mathcal{O}[\mathcal{J}_{\kappa}^{\alpha}\mathcal{J}_{\kappa}^{\beta}]$ where the second term is the rest of the expansion. Thus, in a general basis  $\mathsf{S}_{\alpha}(\zeta)\mathscr{H}_{\kappa}^{\beta}(\zeta)\mathsf{S}_{\alpha}^{\dagger}(\zeta)\approx\mathscr{H}_{\kappa}^{\beta}(\zeta)$ and in accordance with (\ref{equ26}) we construct
\begin{eqnarray}\label{equ28}
\mathbf{U}_{\kappa}(\tau_{\kappa}, \tau_{\kappa-1})\approx
	\mathsf{S}_{-}(\tau_{\kappa},\tau_{\kappa-1})\mathsf{S}_{0}(\tau_{\kappa},\tau_{\kappa-1})\mathsf{S}_{+}(\tau_{\kappa},\tau_{\kappa-1}),
\end{eqnarray}
where
\begin{eqnarray}\label{equ29}
\mathsf{S}_{\beta}(\tau_{\kappa}, \tau_{\kappa-1})=
\left[ {\begin{array}{*{20}c}
	a_{\kappa}^{\beta} & b_{\kappa}^{\beta}e^{i\Psi_{\kappa}^{\beta}}\\
	-b_{\kappa}^{\beta*}e^{-i\Psi_{\kappa}^{\beta}} & a_{\kappa}^{\beta*}
	\end{array} } \right],
\end{eqnarray} 
and where
\begin{eqnarray}\label{equ30a}
a_{\kappa}^{\beta}=
-\frac{i\Gamma(i\delta_{\kappa}^{\beta}+1)}{\sqrt{2\pi}}\Big[D_{-i\delta_{\kappa}^{\beta}}(-iz_{\kappa}^{\beta})D_{-i\delta_{\kappa}^{\beta}-1}(iz_{\kappa-1}^{\beta})+D_{-i\delta_{\kappa}^{\beta}}(iz_{\kappa}^{\beta})D_{-i\delta_{\kappa}^{\beta}-1}(-iz_{\kappa-1}^{\beta})\Big],
\end{eqnarray}
and
\begin{eqnarray}\label{equ30b}
b_{\kappa}^{\beta}=
\frac{\Gamma(i\delta_{\kappa}^{\beta}+1)e^{-i\pi/4}}{\sqrt{2\pi\delta_{\kappa}^{\beta}}}\Big[D_{-i\delta_{\kappa}^{\beta}}(-iz_{\kappa}^{\beta})D_{-i\delta_{\kappa}^{\beta}}(iz_{\kappa-1}^{\beta})-D_{-i\delta_{\kappa}^{\beta}}(iz_{\kappa}^{\beta})D_{-i\delta_{\kappa}^{\beta}}(-iz_{\kappa-1}^{\beta})\Big],
\end{eqnarray}
are  $SU(2)$ Caley-Klein parameters defined such that $|a_{\kappa}^{\beta}|^{2}+|b_{\kappa}^{\beta}|^{2}=1$. Here, $
\delta_{\kappa}^{\beta}=\mathcal{J}_{\kappa}^{\beta}(A/\omega)\mathcal{J}_{\kappa}^{\beta}(A/\omega)
$ and $z_{\kappa}^{\beta}(\tau)=(\tau_{\kappa}+\omega_{\kappa}^{\beta}) e^{-i\pi/4}$.  Returning to Eq.(\ref{equ28})  equipped with (\ref{equ29}), we compute
\begin{eqnarray}\label{equ31}
\mathbf{U}_{\kappa}(\tau_{\kappa}, \tau_{\kappa-1})
\approx
\left[ {\begin{array}{*{20}c}
	c_{\kappa} & d_{\kappa}\\
	-d_{\kappa}^{*} & c_{\kappa}^{*}
	\end{array} } \right].
\end{eqnarray}
Here,
 \begin{eqnarray}\label{equ32a}
 c_{\kappa}=a_{\kappa}^{+} \Big(a_{\kappa}^{0} a_{\kappa}^{-} - b_{\kappa}^{0*} b_{\kappa}^{-}e^{-i[\Psi_{\kappa}^{0}-\Psi_{\kappa}^{-}]}\Big) - 
 b_{\kappa}^{+*} \Big(a_{\kappa}^{-} b_{\kappa}^{0}e^{i[\Psi_{\kappa}^{0}-\Psi_{\kappa}^{+}]} + 
 a_{\kappa}^{0*} b_{\kappa}^{-}e^{i[\Psi_{\kappa}^{-}-\Psi_{\kappa}^{+}]}\Big),
 \end{eqnarray}
 and
 \begin{eqnarray}\label{equ32b}
 d_{\kappa}=b_{\kappa}^{-}e^{i\Psi_{\kappa}^{-}}\Big(a_{\kappa}^{0*} a_{\kappa}^{+*} - b_{\kappa}^{0*} b_{\kappa}^{+}e^{-i[\Psi_{\kappa}^{0}-\Psi_{\kappa}^{+}]}\Big) + 
 a_{\kappa}^{-} \Big(a_{\kappa}^{+*} b_{\kappa}^{0}e^{i\Psi_{\kappa}^{0}} + a_{\kappa}^{0} b_{\kappa}^{+}e^{i\Psi_{\kappa}^{+}}\Big).
 \end{eqnarray}
 As the ratio $A/\omega\ll1$, the main contribution in the product of propagators (\ref{equ31}) and the series expansion of Bessel functions comes from $\kappa=0$. The desired populations are given by $P_{\uparrow\to\uparrow}(+\infty)=|\bra{\uparrow}\mathbf{U}_{\kappa=0}(+\infty,-\infty)\ket{\uparrow}|^2$ for the population remained and $P_{\uparrow\to\downarrow}(+\infty)=|\bra{\uparrow}\mathbf{U}_{\kappa=0}(+\infty,-\infty)\ket{\downarrow}|^2$ for the population transferred. We confirm this with the help of a numerical text (see Fig.\ref{Figure2a}) implemented by considering the large time asymptotic $\tau=+\infty$ of equations (\ref{equ30a}) and (\ref{equ30b}) i.e.
  \begin{eqnarray}\label{equ33}
   a_{\kappa=0}^{\beta}=\sqrt{\exp(-2\pi\delta_{\kappa=0}^{\beta})},\quad {\rm and} \quad b_{\kappa=0}^{\beta}=\sqrt{1-\exp(-2\pi\delta_{\kappa=0}^{\beta})}\exp(-i\chi_{\kappa=0}^{\beta}), 
   \end{eqnarray}
   where 
   $\chi_{\kappa}^{\beta}=\pi/4+\arg\Gamma(1-i \delta_{\kappa}^{\beta})+\delta_{\kappa}^{\beta} (\log \delta_{\kappa}^{\beta}-1)$ is the Stokes phase originating from the large series expansion of Weber's functions. Our analytical solutions read
     \begin{eqnarray}\label{equ33a}
     P_{\uparrow\to\uparrow}(\infty)=\sum_{j,j'=1}^{4}\mathsf{P}_{j}\mathsf{P}_{j'}\cos[\xi_{j}-\xi_{j'}],
     \end{eqnarray}
     and $P_{\uparrow\to\downarrow}(\infty)=1-P_{\uparrow\to\uparrow}(\infty)$ with the functions (different from probabilities)
     \begin{eqnarray}\label{equ33b}
     \mathsf{P}_1=\Big|a^{-}_{\kappa=0}a^{0}_{\kappa=0}a^{+}_{\kappa=0}\Big|,\quad \mathsf{P}_2=-\Big|b^{-}_{\kappa=0}b^{0}_{\kappa=0}a^{+}_{\kappa=0}\Big|, \quad \mathsf{P}_3=-\Big|a^{-}_{\kappa=0}b^{0}_{\kappa=0}b^{+}_{\kappa=0}\Big|,\quad \mathsf{P}_4=-\Big|b^{-}_{\kappa=0}a^{0}_{\kappa=0}b^{+}_{\kappa=0}\Big|,
     \end{eqnarray}
     and
     \begin{eqnarray}\label{equ33c}
     \nonumber  \xi_1=0, \quad \xi_2=[\Psi_{\kappa=0}^{0}-\Psi_{\kappa=0}^{-}]-[\chi_{\kappa=0}^{0}-\chi_{\kappa=0}^{-}], \quad \hspace{2cm}\\\xi_3=[\Psi_{\kappa=0}^{0}-\Psi_{\kappa=0}^{+}]-[\chi_{\kappa=0}^{0}-\chi_{\kappa=0}^{+}],\quad \xi_4=[\Psi_{\kappa=0}^{-}-\Psi_{\kappa=0}^{+}]-[\chi_{\kappa=0}^{-}-\chi_{\kappa=0}^{+}].
     \end{eqnarray}
In general in this case, $a_{\kappa=0}^{+}=a_{\kappa=0}^{-}$ and $b_{\kappa=0}^{+}=b_{\kappa=0}^{-}$. These results hold in the limits $\Delta_{\alpha}/\sqrt{v}\ll1$ and $A/\omega\ll1$ for arbitrary $\varepsilon_{0}$ and $\phi$. Important remark, for application of these formula, we only require the RF field to be tuned such that $A\ll\omega$ to guarantee that $A/\omega\ll1$; in other words $A/\sqrt{v}$ and $\omega/\sqrt{v}$ can be arbitrarily chosen as long as the condition of validity $A/\omega\ll1$ is satisfied. For example on Fig.\ref{Figure2a} upper panel, $A/\sqrt{v}=1$ and  $\omega/\sqrt{v}=50$ and on the lower panel $A/\sqrt{v}=29$ and  $\omega/\sqrt{v}=100$. These two cases fall in the range of validity required. This is confirmed by the satisfactory agreement observed between analytical and numerical results. The Schr\"odinger picture investigated here is relevant to describe population dynamics and population transfer in the spin qubit. To perform coding and reading out of information, another complementary picture is more appropriate: the Bloch's picture.

\section{Bloch picture}\label{Sec4}
Given that the wave function $|\Phi (\tau)\rangle$ in the Schr\"odinger picture has symmetry operations that belong to the finite dimensional Lie group $SU(2)$ and given that $SU(2)/C_{2}$ (where $C_{2}$ is the permutation group of two objects) doubly covers $SO(3)$ (group of rotations and angular momentum), the local isomorphism $SO(3)\approx SU(2)\times SU(2)$ allows us to define the object 
$\boldsymbol{\mathrm{\rho}}(\tau)=|\Phi (\tau)\rangle\langle\Phi (\tau)|$, known as density matrix (DM) with $SO(3)$ symmetry and satisfying the von Neumann equation $i\dot{\boldsymbol{\mathrm{\rho}}}(\tau)=[\mathcal{H}(\tau), \boldsymbol{\mathrm{\rho}}(\tau)]$. In the diabatic basis, $\boldsymbol{\mathrm{\rho}}(\tau)=\sum_{n,m=1}^{2}\rho_{nm}(\tau)|n\rangle\langle m|$ and 
\begin{equation} \label{eq1}
i\dot{\rho}_{nm}(\tau)=\sum_{\kappa=1}^{2}\Big(\mathcal{H}_{n\kappa}(\tau)\rho_{\kappa m}(\tau)-\mathcal{H}_{m\kappa}(\tau)\rho_{n\kappa}(\tau)\Big),
\end{equation} 
 where $\mathcal{H}_{n\kappa}(\tau)$ are matrix elements of $\mathcal{H}(\tau)$. This equation describes a set of four coupled equations: two for populations $\rho_{11}(\tau)$, $\rho_{22}(\tau)$ (diagonal elements) and two for spin polarizations $\rho_{12}(\tau)$, $\rho_{21}(\tau)$ (off-diagonal elements). The index $\kappa=1,2$ indicate the crossing levels $\ket{\uparrow}$ and $\ket{\downarrow}$. In general, in the absence of decay and/or dissipation $\rho_{\kappa\kappa}(\tau)=\rho_{\kappa\kappa}^{*}(\tau)$ and $\rho_{\kappa\kappa'}(\tau)=\rho_{\kappa'\kappa}^{*}(\tau)$. This allows us to move to a more convenient space suitable for describing qubit dynamics: the Bloch space. We introduce the Bloch vector as $\vec{\mathbf{u}}(\tau)=[2{\rm Re}\rho_{12}(\tau),2{\rm Im}\rho_{21}(\tau), \rho_{11}(\tau)-\rho_{22}(\tau)]$ and relate it to the DM as $\boldsymbol{\mathrm{\rho}}(\tau)=(\mathbf{1}+\vec{\boldsymbol{\mathrm{\sigma}}}\cdot\vec{\mathbf{u}}(\tau))/2$. This representation ensures that if $|\vec{\mathbf{u}}(\tau)|=1$ then ${\rm Tr}\boldsymbol{\mathrm{\rho}}^{2}(\tau)=1$ for pure states. Hereby, the equation for the components of the Bloch vector $u_{\alpha}={\rm Tr}(\boldsymbol{\mathrm{\sigma}}_{\alpha}\boldsymbol{\mathrm{\rho}})$ (with $\alpha=x,y,z$) is obtained as 
 \begin{eqnarray}\label{eq2a}
 \frac{du_{\alpha}}{d\tau}=\sum_{\beta,\gamma}\epsilon_{\alpha\beta\gamma}b_{\beta}u_{\gamma},
 \end{eqnarray}
 (Here, $\beta,\gamma=x,y,z$) after considering ${\rm Tr}(\boldsymbol{\mathrm{\sigma}}_{\alpha}\boldsymbol{\mathrm{\sigma}}_{\beta})=2\delta_{\alpha\beta}$ (where $\delta_{\alpha\beta}$ is the Kronecker Delta symbol which takes value $1$ when $\alpha=\beta$ and zero otherwise) and the fact that the single Casimir operator $\vec{\boldsymbol{\mathrm{\sigma}}}^{2}$ commutes with all the generators $\boldsymbol{\mathrm{\sigma}}_{\alpha}$ of the $su(2)$ algebra. From here, using the properties of the fully anti-symmetric Levi-Civita symbol, one can show that the dynamics of the Bloch vector is equivalent to the classical motion of magnetic moment in a magnetic field i.e. $\dot{\vec{\mathbf{u}}}(\tau)=-\vec{\mathbf{u}}\wedge \vec{ b}(\tau)$ (where the overhead dot denotes time derivative). If we assume $u_{\alpha}(\tau)=\langle \boldsymbol{\mathrm{\sigma}}_{\alpha}(\tau)\rangle$, then the average of the spin vector $\langle\vec{S}(\tau)\rangle=\langle\vec{\boldsymbol{\mathrm{\sigma}}}(\tau)\rangle/2$ and a pure state $|\Phi \rangle=\cos[\vartheta_{\rm az}/2]\exp[-i\vartheta_{\rm pol}/2]\ket{\uparrow}+\sin[\vartheta_{\rm az}/2]\exp[i\vartheta_{\rm pol}/2]\ket{\downarrow}$ is represented on the surface of the Bloch sphere through the Bloch vector $\vec{\mathbf{u}}(\tau)=[\sin\vartheta_{\rm az}\cos\vartheta_{\rm pol}, \sin\vartheta_{\rm az}\sin\vartheta_{\rm pol},  \cos\vartheta_{\rm az}]$ providing the azimuthal and polar angles $\vartheta_{\rm az}=\arcsin[u_{x}^2+u_{y}^2]^{1/2}$ and $\vartheta_{\rm pol}=\arctan(u_{y}/u_{x})$ respectively (an example is shown in Fig.\ref{Figure1} for the standard LZSM problem) with $0\leqslant\vartheta_{\rm pol}\leqslant 2\pi$ and $0\leqslant\vartheta_{\rm az}\leqslant \pi$. A spin flip is achieved at $\tau_{\rm flip}$ when the various fields are tuned such that the Bloch vector leaves the north pole and migrate to the south pole by sweeping an angle $\vartheta_{\rm az}(\tau_{\rm flip})=\pi$ and in this case $u_z(\tau_{\rm flip})=-1$. 
 
\begin{figure}[]
	\centering
	\begin{center} 
		\includegraphics[width=6.3cm, height=5.6cm]{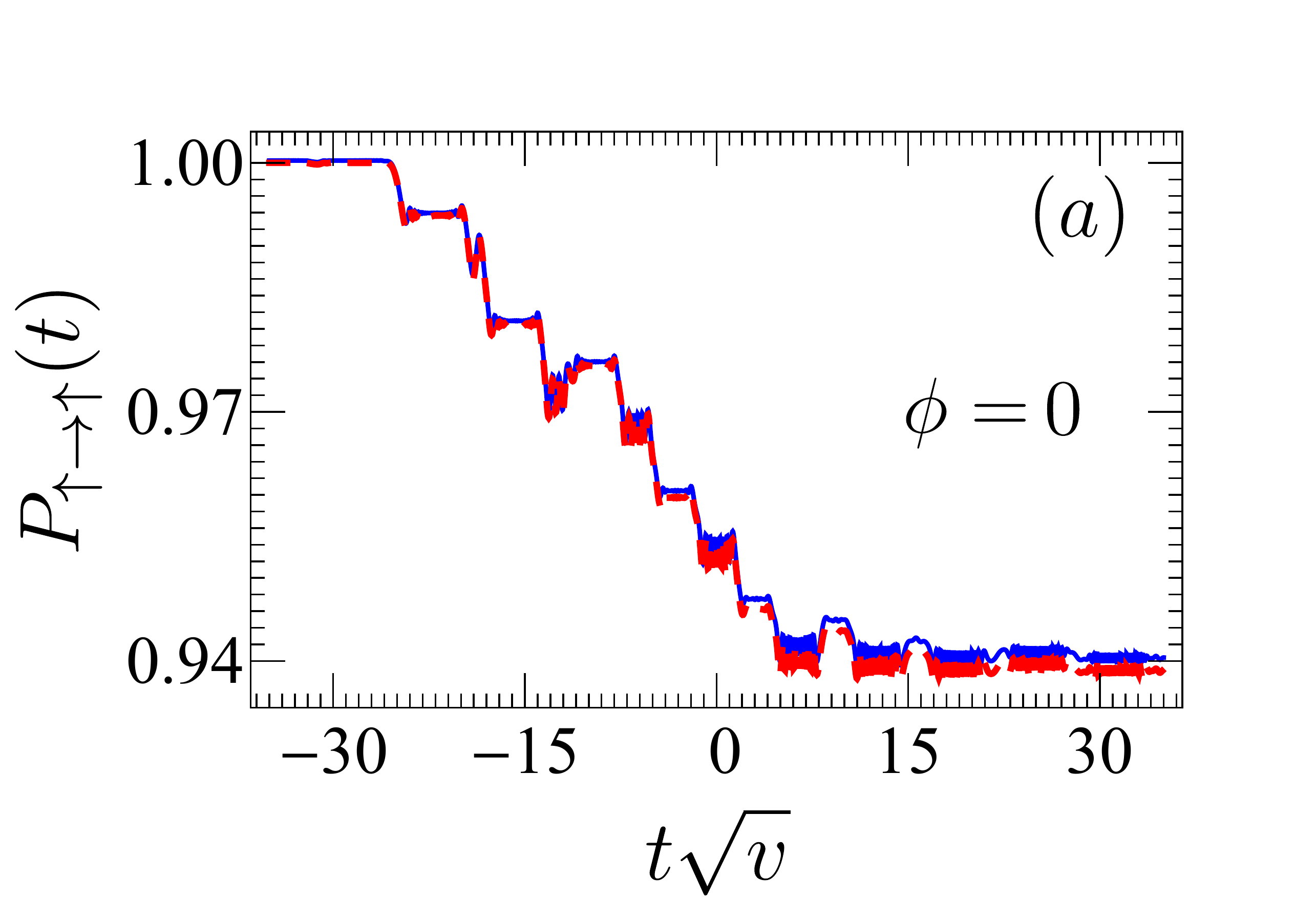}\hspace{-0.7cm}
		\includegraphics[width=6.3cm, height=5.6cm]{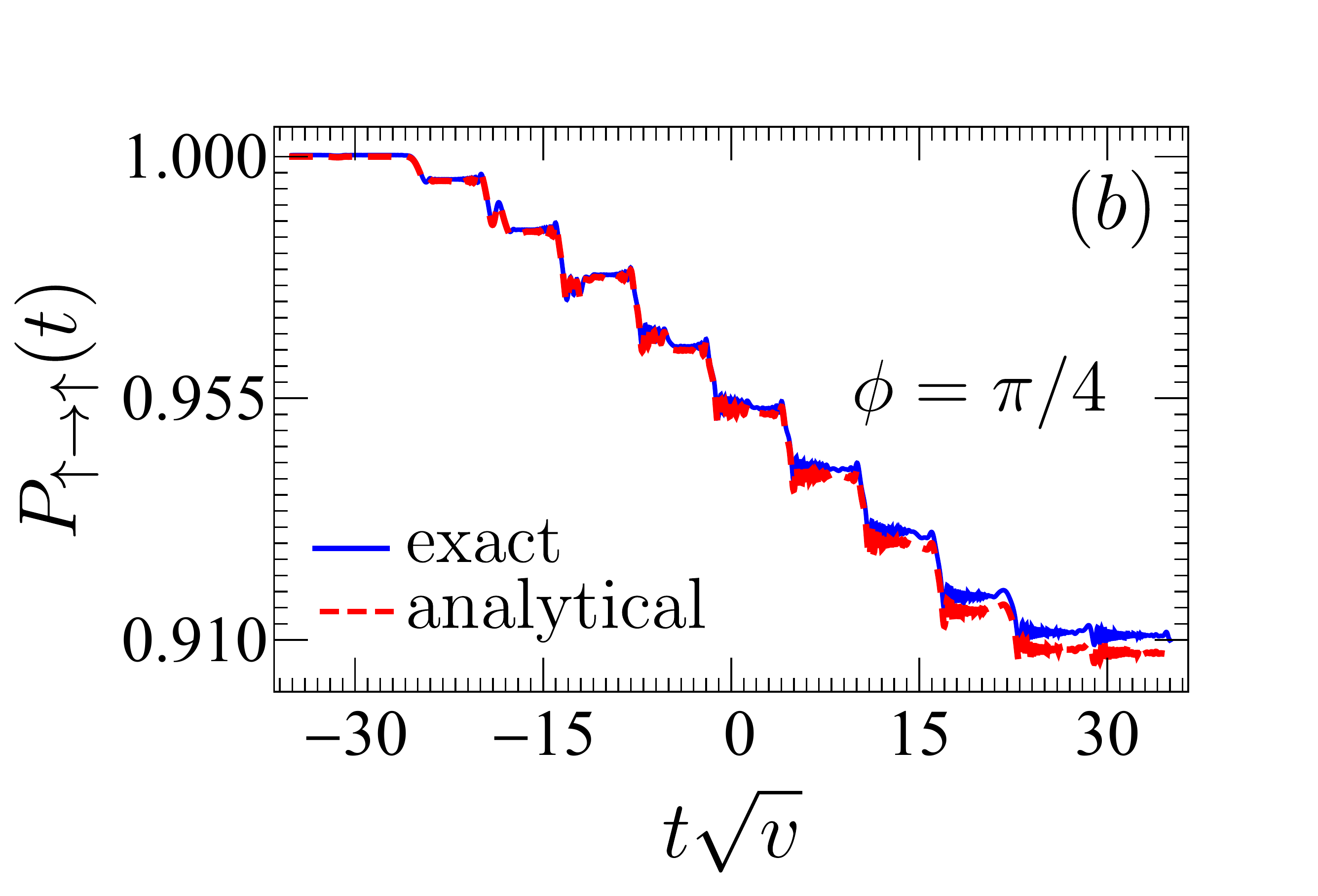}\hspace{-0.7cm}
		\includegraphics[width=6.3cm, height=5.6cm]{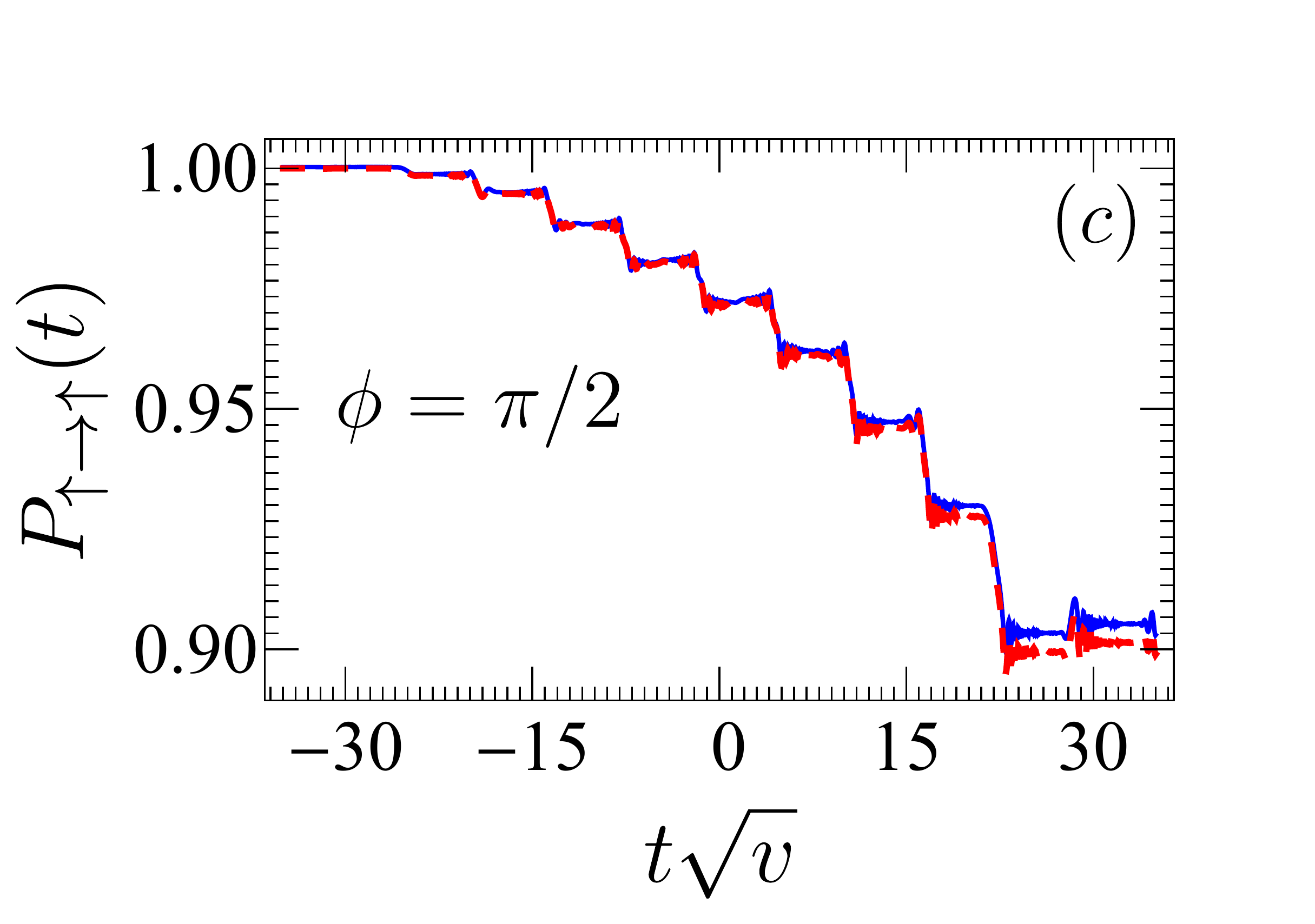}\\\vspace{-0.8cm}
		\includegraphics[width=6.3cm, height=5.6cm]{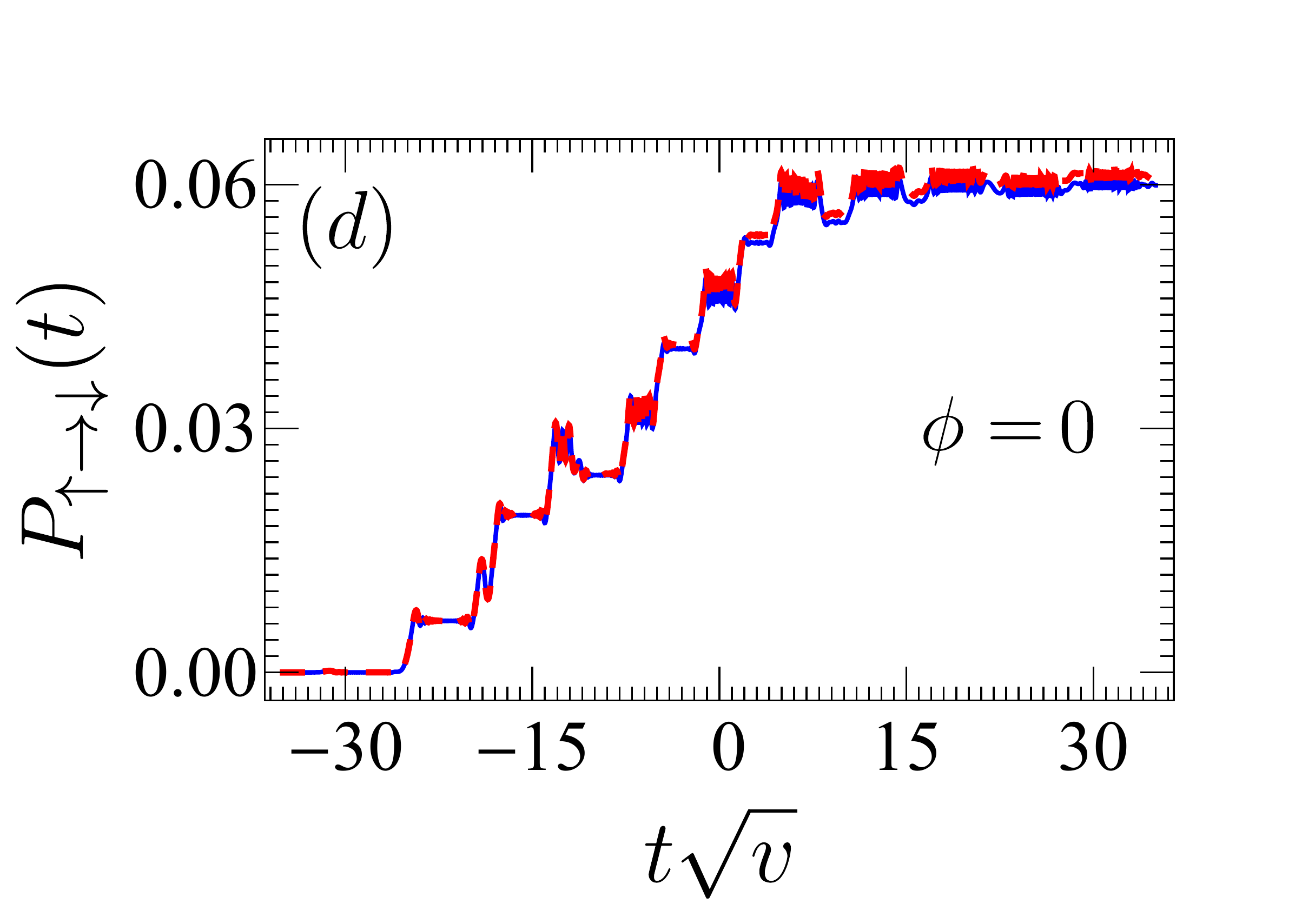}\hspace{-0.7cm}
		\includegraphics[width=6.3cm, height=5.6cm]{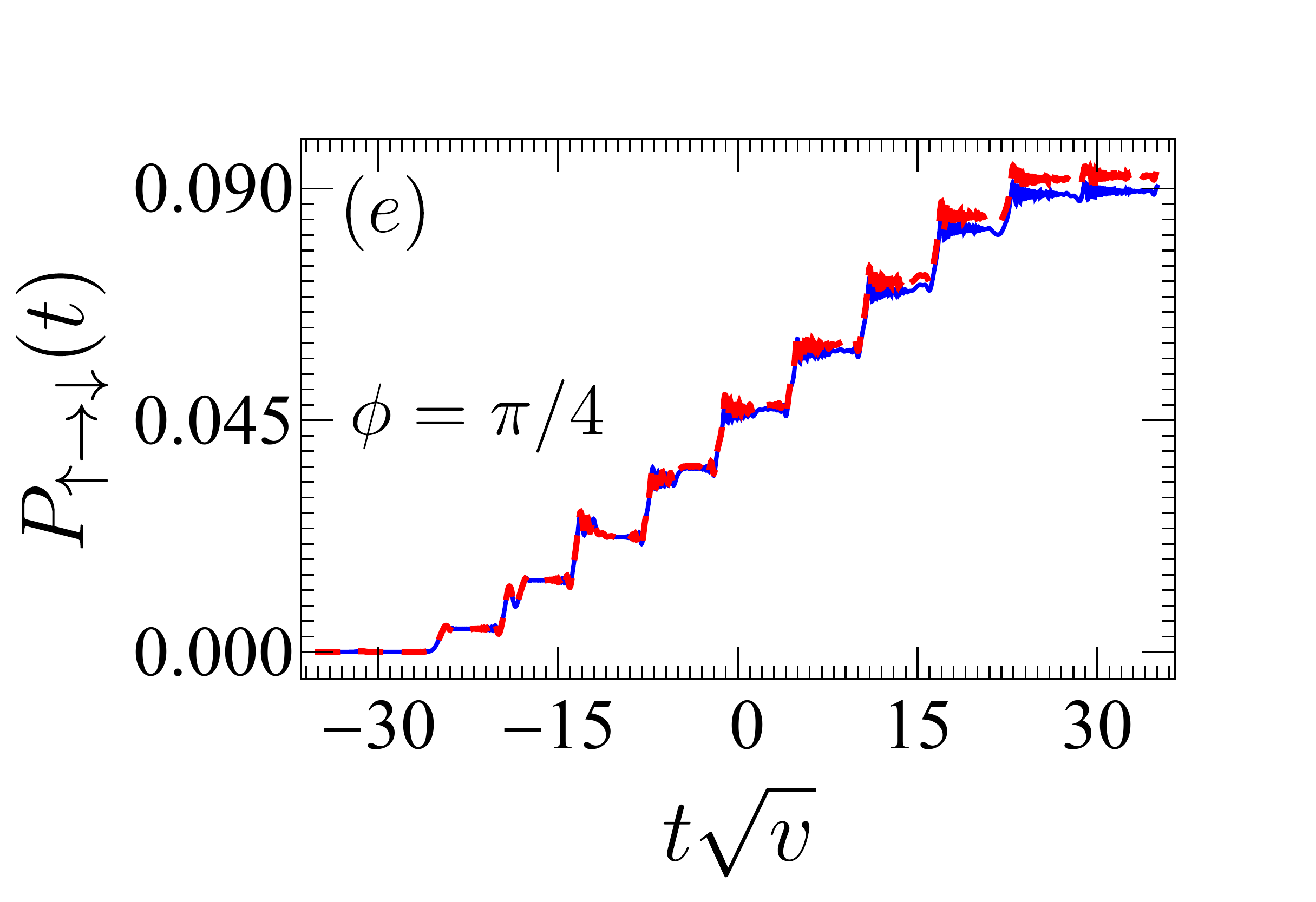}\hspace{-0.7cm}
		\includegraphics[width=6.3cm, height=5.6cm]{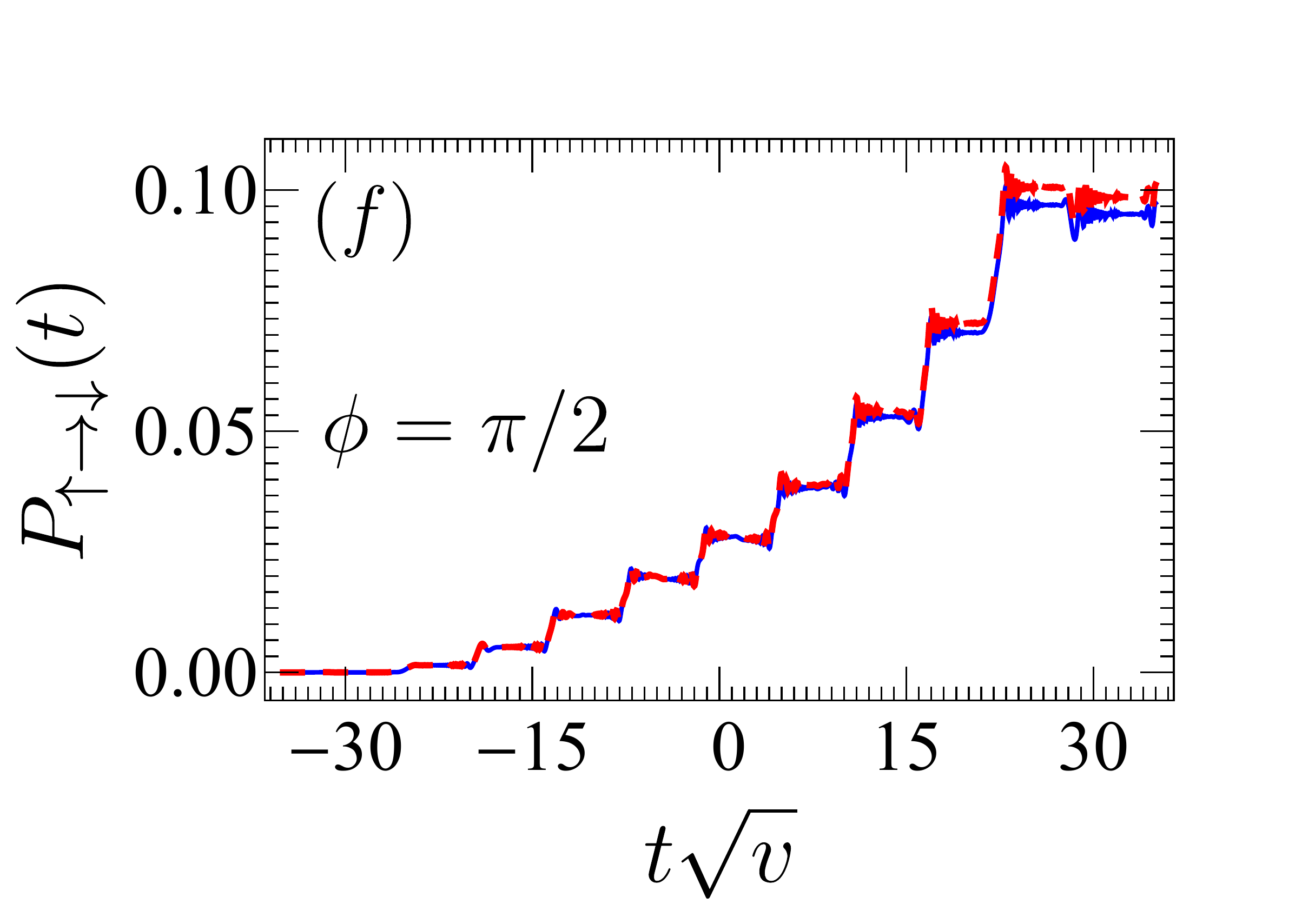} 
		\vspace{-0.5cm}
		\caption{ Cascaded LZSM transitions. Comparison between analytical (\ref{equ4.3}) and numerical solutions of the von Neumann equation (\ref{eq2a}). For calculations, $A/\sqrt{v}=25$, $\varepsilon_0/\sqrt{v}=0.5$, $\mathcal{A}_{f}/\sqrt{v}=0.08$, $\omega=\omega_{f}=\sqrt{v}$ and $\Delta/\sqrt{v}=0.07$. The indexes $n$ and $m$ are truncated to $40$ such that the analytical results match the numerical solutions. Blue solid lines are exact numerical results while red dashed lines are analytical results. We observe a remarkable agreement between analytical and numerical results in the said limit given that both results are barely discernible.}
		\label{Figure3}
	\end{center}
\end{figure}

\begin{figure}[]
 	\centering
 	\begin{center} 
 		\includegraphics[width=6.3cm, height=5.6cm]{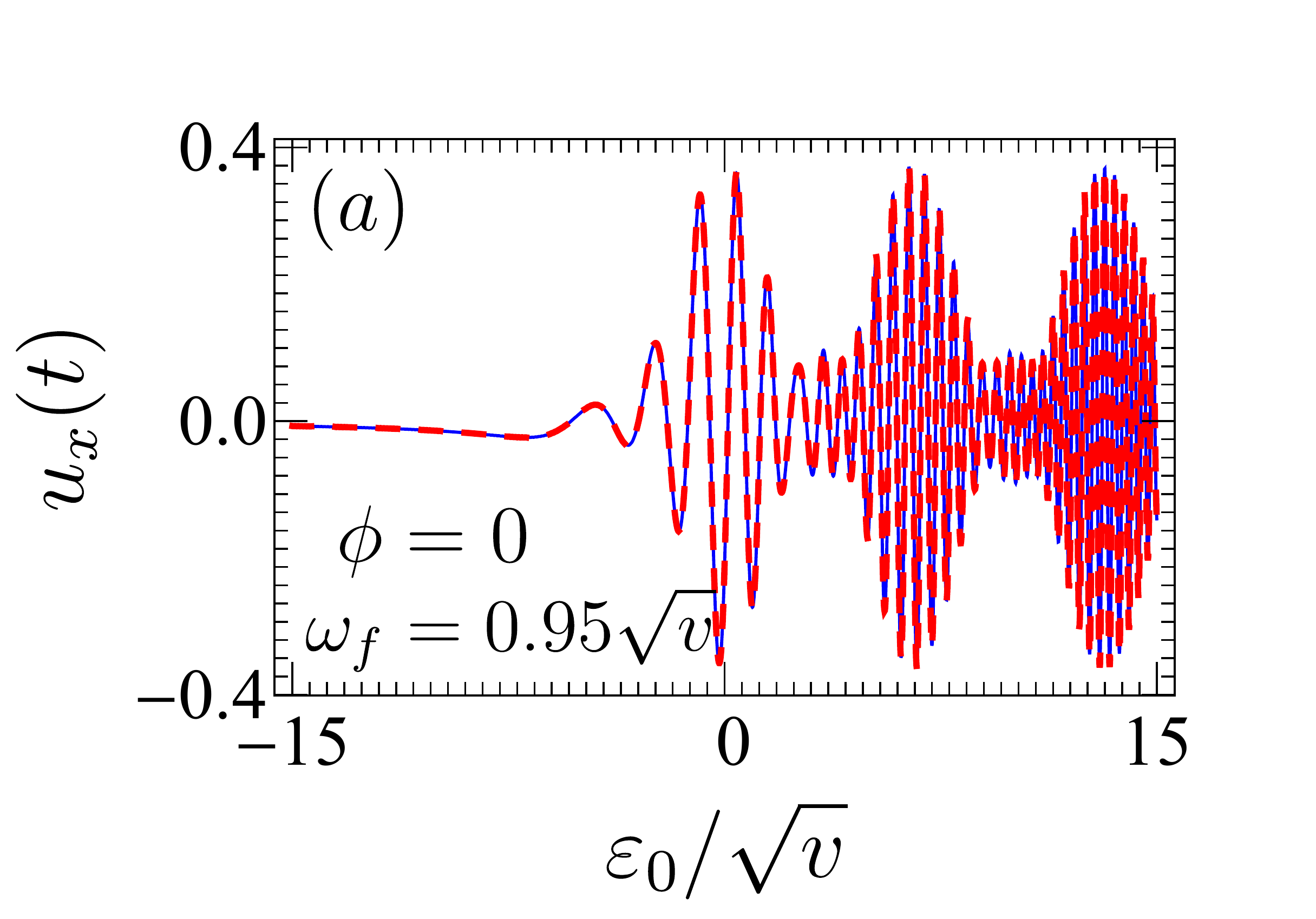}\hspace{-0.7cm}
 		\includegraphics[width=6.3cm, height=5.6cm]{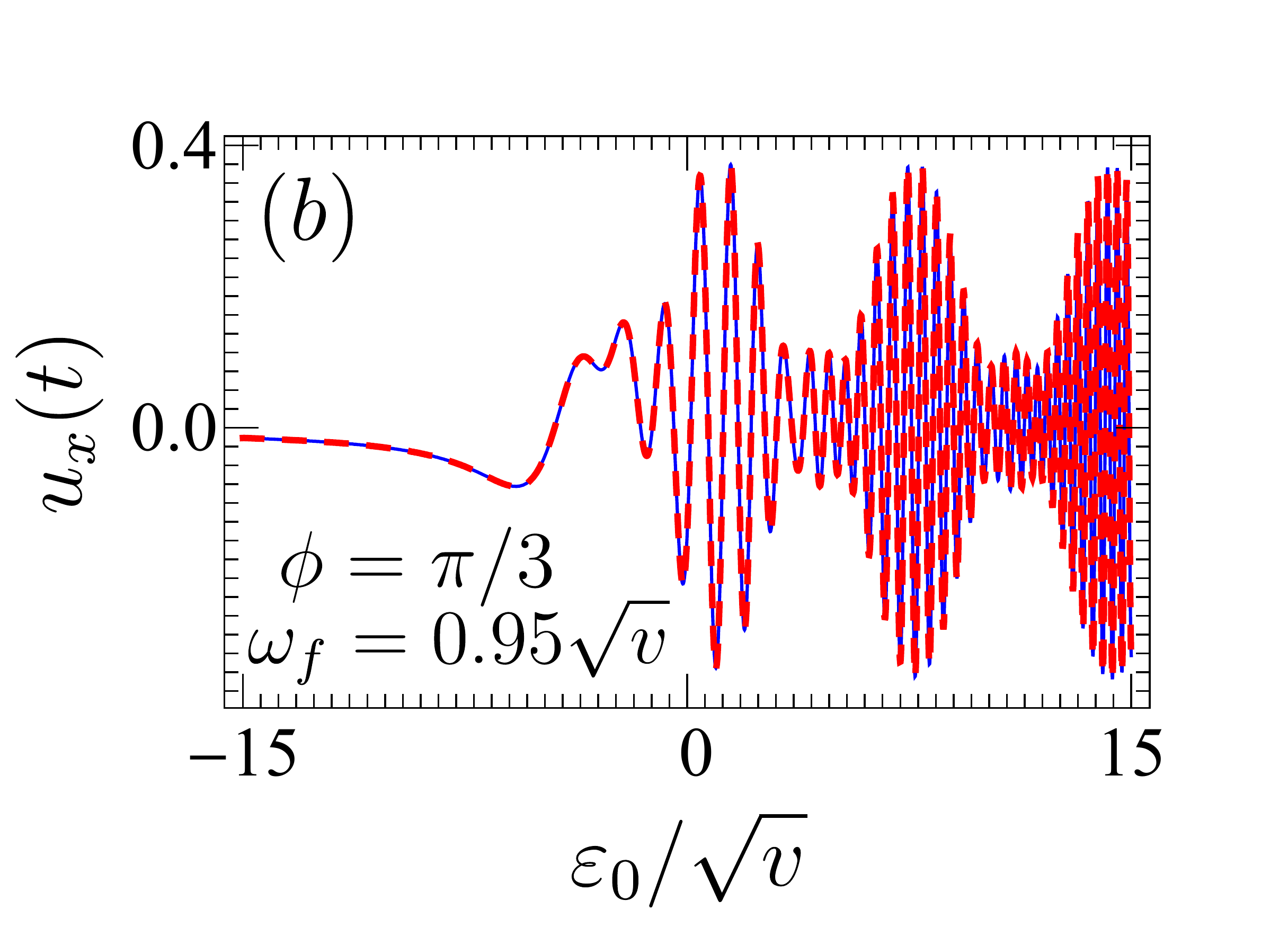}\hspace{-0.7cm}
 		\includegraphics[width=6.3cm, height=5.6cm]{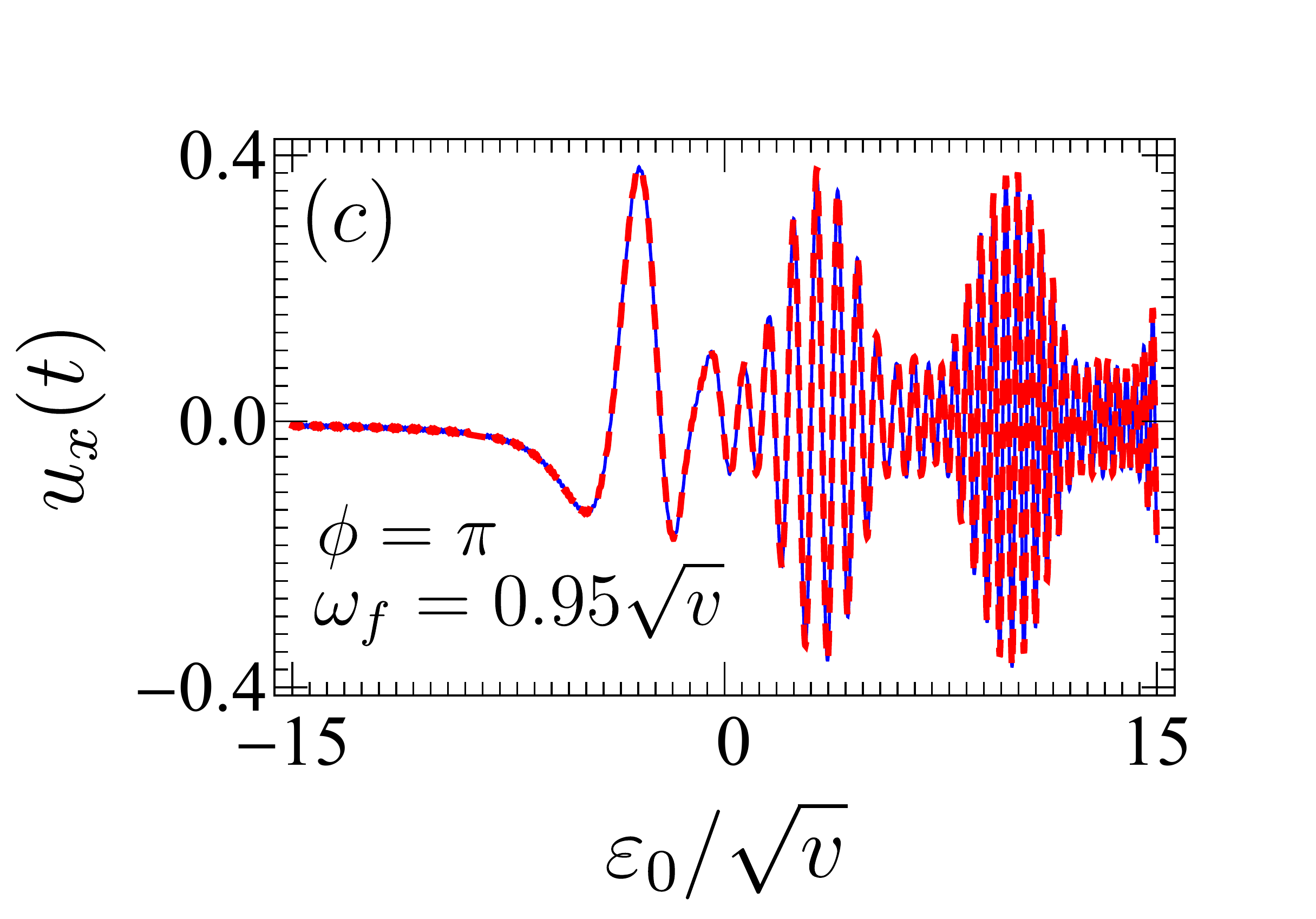}\\\vspace{-0.9cm}
 		\includegraphics[width=6.3cm, height=5.6cm]{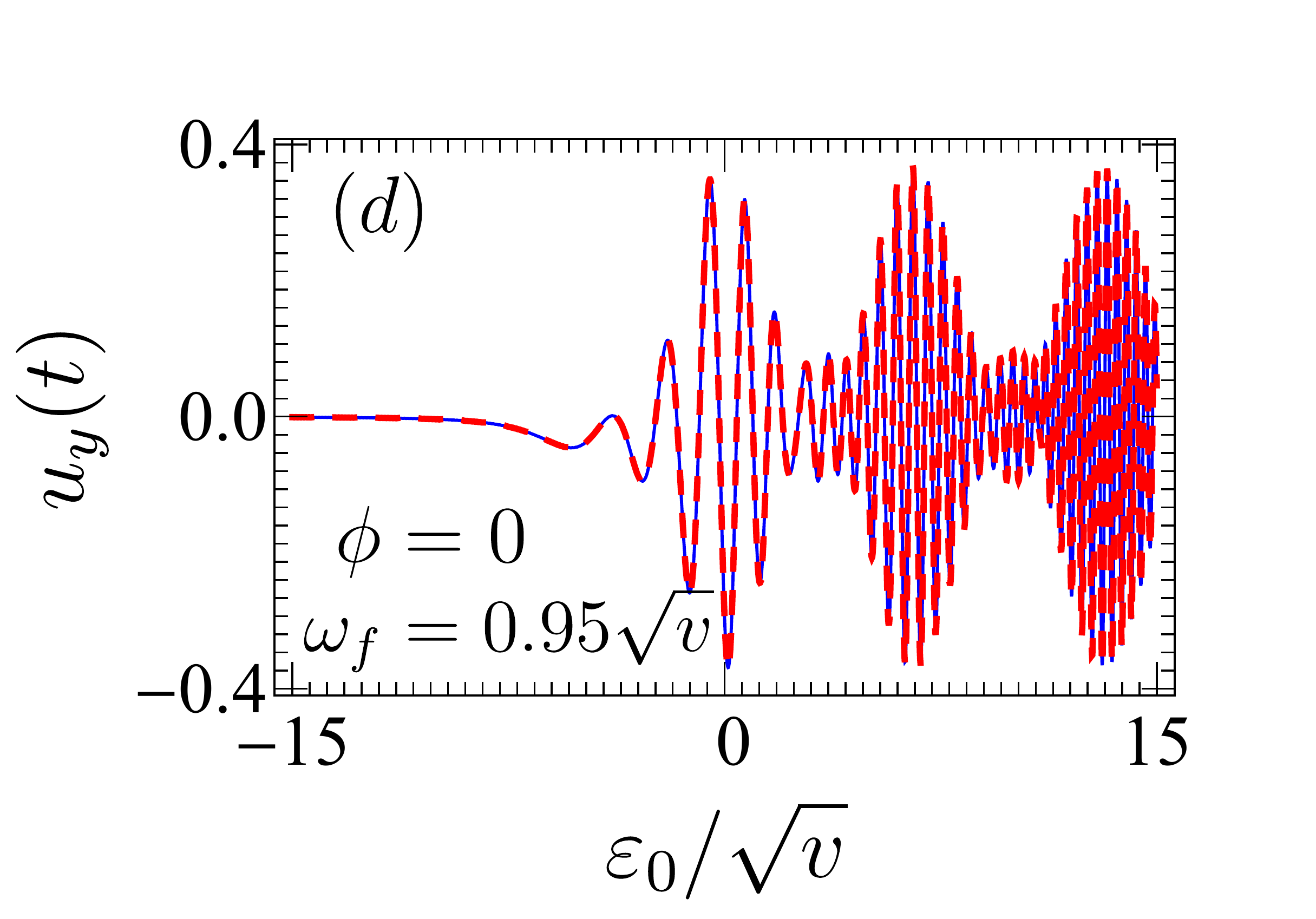}\hspace{-0.7cm}
 		\includegraphics[width=6.3cm, height=5.6cm]{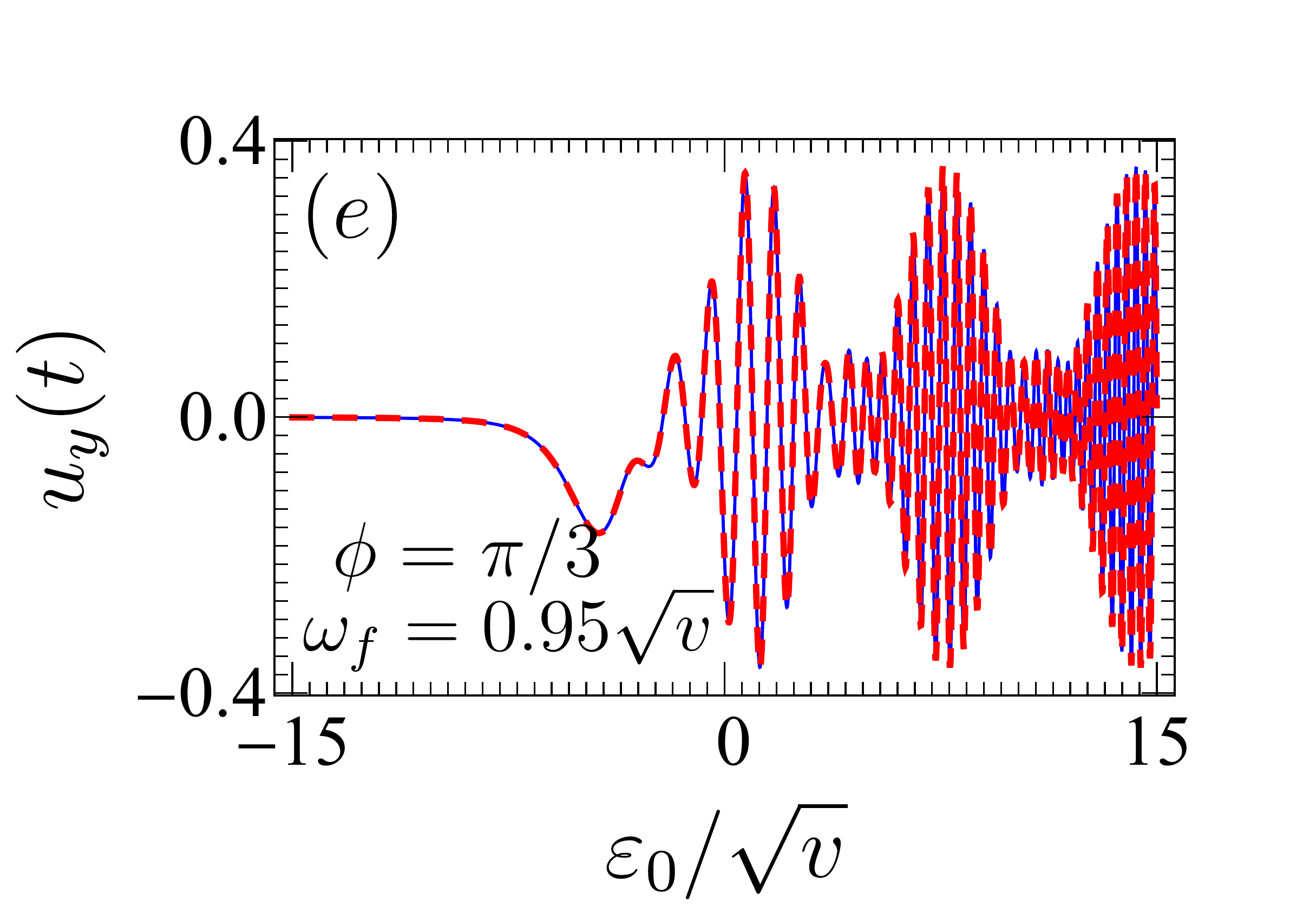}\hspace{-0.7cm}
 		\includegraphics[width=6.3cm, height=5.6cm]{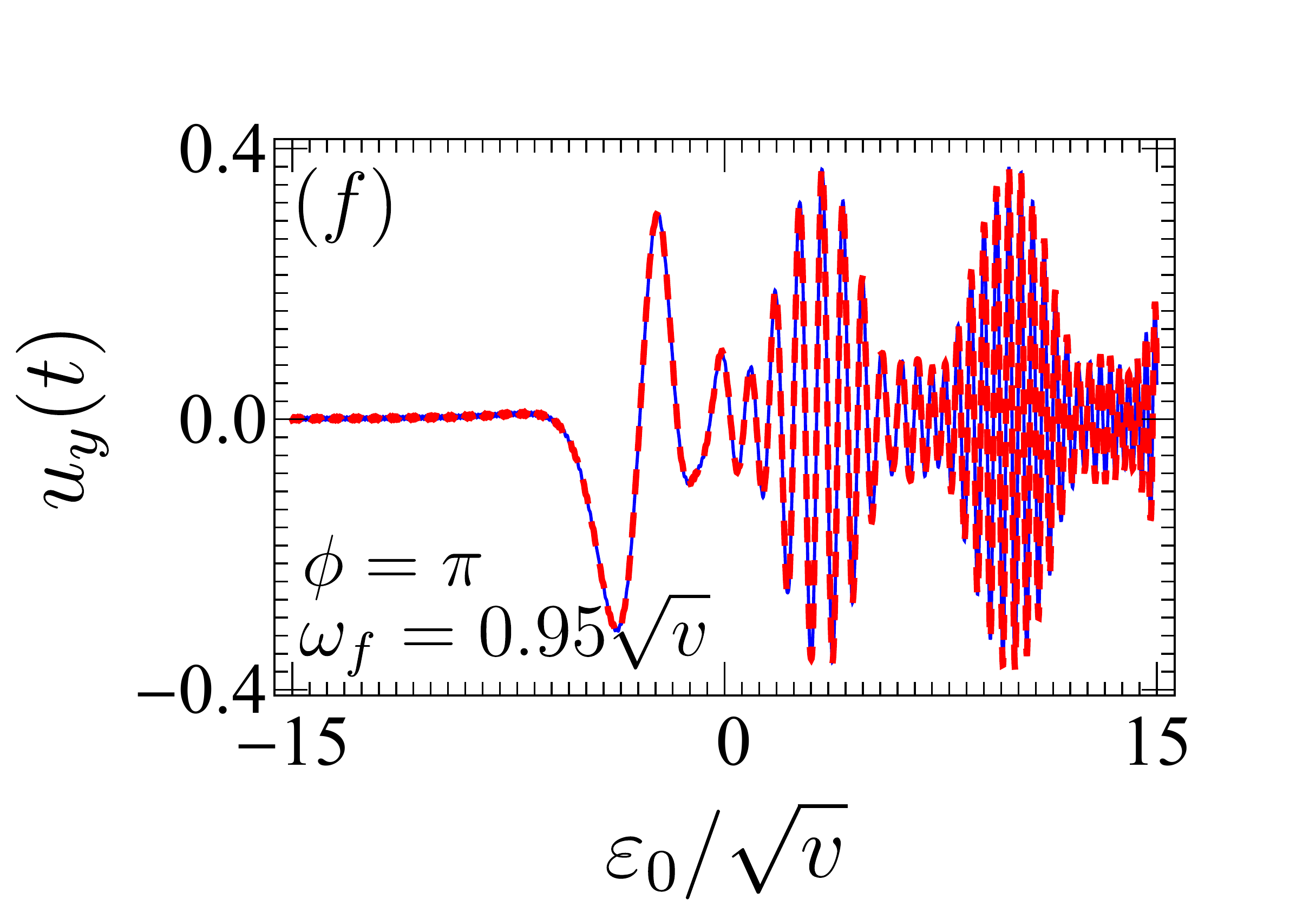}
 		\vspace{-0.5cm}
 		\caption{ Comparison between analytical (solid blue lines) and numerical results (red dashed lines) for spin polarizations. For calculations, $A/\sqrt{v}=0.05$, $\mathcal{A}_{f}/\sqrt{v}=0.085$, $\omega=\sqrt{v}$ and $\Delta/\sqrt{v}=0.07$. The indexes $n$ and $m$ are truncated to $40$ such that the analytical results match the numerical solutions. The integration time runs from $t\sqrt{v}=-50$ to $t\sqrt{v}=50$. It is barely impossible to distinguish between both solutions.}
 		\label{Figure6}
 	\end{center}
 \end{figure}

 After eliminating all coherence factors from the equation for population difference, one achieves
\begin{eqnarray}\label{eq2aa}
u_{x}(\tau)=2\sum_{n}J_{n}\Big(\frac{A}{\omega}\Big)\sum_{\beta,m}\mathcal{J}_{m}^{\beta}\Big(\frac{A}{\omega}\Big)\int_{-\infty}^{\tau}d\tau_{1}\sin\Big[\mathcal{K}_{n}^{0}(\tau)-\mathcal{K}_{m}^{\beta}(\tau_{1})\Big]u_{z}(\tau_{1}),
\end{eqnarray}
\begin{eqnarray}\label{eq2ab}
u_{y}(\tau)=-2\sum_{n}J_{n}\Big(\frac{A}{\omega}\Big)\sum_{\beta,m}\mathcal{J}_{m}^{\beta}\Big(\frac{A}{\omega}\Big)\int_{-\infty}^{\tau}d\tau_{1}\cos\Big[\mathcal{K}_{n}^{0}(\tau)-\mathcal{K}_{m}^{\beta}(\tau_{1})\Big]u_{z}(\tau_{1}),
\end{eqnarray}
and
\begin{eqnarray}\label{eq2}
u_{z}(\tau)=1-\sum_{n,m}\sum_{\alpha,\beta}\mathcal{J}_{n}^{\alpha}\Big(\frac{A}{\omega}\Big)\mathcal{J}_{m}^{\beta}\Big(\frac{A}{\omega}\Big)\int_{-\infty}^{\tau}d\tau_{1}\int_{-\infty}^{\tau_{1}}d\tau_{2}\cos\Big[\mathcal{K}_{n}^{\alpha}(\tau_{1})-\mathcal{K}_{m}^{\beta}(\tau_{2})\Big]u_{z}(\tau_{2}),
\end{eqnarray}
where 
\begin{eqnarray}\label{eq3}
\mathcal{K}_{n}^{\alpha}(\tau)=\frac{1}{2}[\tau+\omega_{n}^{\alpha}]^{2}-\Psi_{n}^{\alpha}.
\end{eqnarray}
(Here, $n,m$ run from $-\infty$ to $+\infty$ while $\alpha,\beta=-,0,+$). 
Equation (\ref{eq2}) is the central equation in this picture given that the desired populations are obtained as $P_{\uparrow\to\uparrow}(\tau)=[1+u_{z}(\tau)]/2$ and  $P_{\uparrow\to\downarrow}(\tau)=[1-u_{z}(\tau)]/2$. Equations (\ref{eq2aa}) and (\ref{eq2ab}) as spin polarizations technically help realizing spin flip and performing coding and reading out of qubit\cite{Muga}. They also help minimizing the time spent by the spin in other directions than the $z$-direction. They may also be employed for controlling and speeding up spin flip (see Ref.\onlinecite{Sinitsyn} for ample discussion). For instance, in order to achieve a full spin flip, it is mandatory to impose by all possible means that $u_y(\tau)=0$. Amongst other things Eq.(\ref{eq2aa}) allows to evaluate the spin flip duration. For the aforementioned reasons, Eqs.(\ref{eq2aa})-(\ref{eq2}) should be solved. On the other hand, due to their actual complexity, they cannot be exactly solved and their solutions written in closed-form. However, the limit $\Delta_{\alpha}^{2}/v\ll1$ provides a solid test-bed for approximating the gross temporal profile of population during non-adiabatic evolutions and evaluation of spin polarizations. Thus, in the non-adiabatic limit $\Delta_{\alpha}^{2}/v\ll1$, Eqs. (\ref{eq2aa})-(\ref{eq2}) are perturbatively solved  considering the  initial conditions $\vec{\mathbf{u}}(-\infty)=[0,0,1]^{T}$ (the Bloch vector is at the north pole of the Bloch sphere). After a long, tedious but straightforward algebra, we obtain 
\begin{eqnarray}\label{eq4.3aa}
u_{x}(\tau)\approx 2\sqrt{\pi}\sum_{n,\alpha}\mathcal{J}_{n}^{\alpha}\Big(\frac{A}{\omega}\Big)\Big(a_{c}(\tau)\mathsf{L}\Big(\tau+\omega_{n}^{\alpha},\Psi_{n}^{\alpha}\Big)+a_{s}(\tau)\mathsf{M}\Big(\tau+\omega_{n}^{\alpha},\Psi_{n}^{\alpha}\Big)\Big)+\mathcal{O}\Big[\Big(\frac{\Delta_{\alpha}\Delta_{\beta}}{v}\Big)^{3}\Big],
\end{eqnarray}
\begin{eqnarray}\label{eq4.3ab}
u_{y}(\tau)\approx 2\sqrt{\pi}\sum_{n,\alpha}\mathcal{J}_{n}^{\alpha}\Big(\frac{A}{\omega}\Big)\Big(a_{s}(\tau)\mathsf{L}\Big(\tau+\omega_{n}^{\alpha},\Psi_{n}^{\alpha}\Big)-a_{c}(\tau)\mathsf{M}\Big(\tau+\omega_{n}^{\alpha},\Psi_{n}^{\alpha}\Big)\Big)+\mathcal{O}\Big[\Big(\frac{\Delta_{\alpha}\Delta_{\beta}}{v}\Big)^{3}\Big],
\end{eqnarray}
\begin{eqnarray}\label{eq4.3}
u_{z}(\tau)\approx 1-2\pi\Big(\Big[\sum_{n,\alpha}\mathcal{J}_{n}^{\alpha}\Big(\frac{A}{\omega}\Big)\mathsf{L}\Big(\tau+\omega_{n}^{\alpha},\Psi_{n}^{\alpha}\Big)\Big]^2+\Big[\sum_{n,\alpha}\mathcal{J}_{n}^{\alpha}\Big(\frac{A}{\omega}\Big)\mathsf{M}\Big(\tau+\omega_{n}^{\alpha},\Psi_{n}^{\alpha}\Big)\Big]^2\Big)+\mathcal{O}\Big[\Big(\frac{\Delta_{\alpha}\Delta_{\beta}}{v}\Big)^{4}\Big],
\end{eqnarray}
where
$ \mathsf{L}(x,y)=\mathcal{C}(x)\sin(y)-\cos(y)\mathcal{S}(x)$ and
$\mathsf{M}(x,y)=\mathcal{C}(x)\cos(y)+\sin(y)\mathcal{S}(x)$ 
and where we have defined $\mathcal{C}(x)=(\frac{1}{2}+C(\frac{x}{\sqrt{\pi}}))$ and  $\mathcal{S}(x)=(\frac{1}{2}+S(\frac{x}{\sqrt{\pi}}))$. Here, $C(...)$ and $S(...)$ are cosine and sine Fresnel integrals respectively\cite{Book}. We have used the properties (\ref{A5}) and (\ref{A6}) and defined $a_{c}(\tau)=\sum_{n}J_{n}(A/\omega)\cos\mathcal{K}_{n}^{0}(\tau)$ and $a_{s}(\tau)=\sum_{n}J_{n}(A/\omega)\sin\mathcal{K}_{n}^{0}(\tau)$. For further relevant purposes, let us note that $a_{c}^2(\tau)+a_{s}^2(\tau)=\sum_{n,m}J_{n}(A/\omega)J_{m}(A/\omega)\cos[\mathcal{K}_{n}^{0}(\tau)-\mathcal{K}_{m}^{0}(\tau)]$ and that $a_{c}^2(\tau)+a_{s}^2(\tau)=1$ when $n=m$ or $n=m=0$ given that $\sum_{n}J_{n}(A/\omega)=1$. In this last case, it is clear that $|\vec{\mathbf{u}}(\tau)|=1$. For nearly-optimal control of qubit, it is relevant to compute the cost functional\cite{Muga, Sinitsyn}. This requires evaluation of the square of spin polarizations. We now wish to open doors for such tasks  for evaluation of $u_{x}^2(\tau)$, $u_{y}^2(\tau)$ and $u_{z}^2(\tau)$. Then, one may need the properties (\ref{A1}) and (\ref{A2}) which for instance simplifies the population difference as
\begin{eqnarray}\label{equ4.3}
u_{z}(\tau)\approx 1-4\pi\sum_{n,m}\sum_{\alpha, \beta}\mathcal{J}_{n}^{\alpha}\Big(\frac{A}{\omega}\Big)\mathcal{J}_{m}^{\beta}\Big(\frac{A}{\omega}\Big)\mathsf{N}^{\alpha, \beta}_{n,m}(\tau)+\mathcal{O}\Big[\Big(\frac{\Delta_{\alpha}\Delta_{\beta}}{v}\Big)^{4}\Big],
\end{eqnarray}
where
\begin{eqnarray}\label{equ4.3a}
\mathsf{N}^{\alpha, \beta}_{n,m}(\tau)=\cos[\Psi_{n}^{\alpha}-\Psi_{m}^{\beta}]F_{+}(\tau+\omega_{n}^{\alpha},\tau+\omega_{m}^{\beta})-\sin[\Psi_{n}^{\alpha}-\Psi_{m}^{\beta}]G_{-}(\tau+\omega_{n}^{\alpha},\tau+\omega_{m}^{\beta})\Big].
\end{eqnarray}
 The functions $F_{\pm}(x,y)$ and $G_{\pm}(x,y)$ are deferred in Appendix \ref{App} (see Eqs.(\ref{equ4.6}) and (\ref{equ4.7}) respectively). The analytic results (\ref{eq4.3aa})-(\ref{eq4.3}) are compared with exact numerical ones obtained by numerically solving the von Newman equation in the non-adiabatic limit (see Figs.\ref{Figure3}-\ref{Figure6}). They hold for arbitrary $\varepsilon_0$, $\omega$, $\omega_{f}$ and $\phi$.  Cascaded LZSM transitions are observed when the amplitude of the RF field widely exceeds the static detuning i.e. $A\gg\varepsilon_0$. This is illustrated in Fig.\ref{Figure3} where simultaneously we observe a remarkable agreement between analytical and numerical data.  For all plots, we have demonstrated that the phase difference $\phi$ between the RF and MW signals alter the coherent dynamic of the qubit. It might be a good parameter for controlling population inversion in the qubit. For instance, on Fig.\ref{Figure3}, the population $P_{\uparrow\to\downarrow}(t)$ rises from $0.06$ to $0.1$ as $\phi$ successively takes the values $0$, $\pi/4$ and $\pi/2$. Remark, at the resonances $\omega_{n}^{\alpha}=\omega_{m}^{\beta}=0$, the index $n$ and $m$ in Eqs.(\ref{eq4.3aa})-(\ref{equ4.3}) respectively take the values $n\equiv n_{\alpha}=-[\varepsilon_0+\alpha\omega_{f}]/\omega$ and $m\equiv m_{\beta}=-[\varepsilon_0+\beta\omega_{f}]/\omega$.  The population difference $u_z(\tau=+\infty)\approx 1-4\pi\delta$ with $\delta$ in Eq.(\ref{equ24}). Considering the probability (\ref{equ22}) in the limit $\mathcal{J}_{n}^{\alpha}(A/\omega)\ll1$, we observe a prefect agreement between the results of the Schr\"odinger and Bloch pictures investigations.  It should also be noted that when the argument of the Bessel function is $A/\omega\ll1$, the dominant contributions in the series of Bessel functions comes from $n=m=0$. This ideally reduces the number of iterations in Eqs.(\ref{eq4.3aa})-(\ref{equ4.3}). When the ratio $A/\omega$ achieves one of the zeros of the Bessel function,  $u_x(\tau)=u_y(\tau)=0$ and $u_z(\tau)=1$: the qubit points  into the $z$-direction. In such a case, there is in addition a coherent destruction of tunneling.

 At large negative time $\tau=-\infty$, the functions in Eqs.(\ref{equ4.6}) and (\ref{equ4.7}) all vanish. The Bloch vector is at the north pole of the Bloch sphere and its amplitude is maximal ($|\vec{\mathbf{u}}(-\infty)|=1$). Ideally, there is a population inversion when the Bloch vector finally ends at the south pole.  At large positive time $\tau=+\infty$, the functions Eqs.(\ref{equ4.6})-(\ref{equ4.7}) simplify and read $F_{+}(+\infty,+\infty)=G_{+}(+\infty,+\infty)=1$ and $F_{-}(+\infty,+\infty)=G_{-}(+\infty,+\infty)=0$. As a direct consequence,
\begin{eqnarray}\label{equ4.8}
u_{z}(+\infty)\approx 1-4\pi\sum_{n,m}\sum_{\alpha, \beta}\mathcal{J}_{n}^{\alpha}\Big(\frac{A}{\omega}\Big)\mathcal{J}_{m}^{\beta}\Big(\frac{A}{\omega}\Big)\cos[\Psi_{n}^{\alpha}-\Psi_{m}^{\beta}]+\mathcal{O}\Big[\Big(\frac{\Delta_{\alpha}\Delta_{\beta}}{v}\Big)^{4}\Big].
\end{eqnarray}
This expression reveals how the spin can be controlled in the quantization direction. It is useful as external clock probe for LZSM tunneling times\cite{Mullen}. Moreover, the results of this section are also relevant for nuclear magnetic resonance (NMR).

\section{More results}\label{Sec5}

In this section, we  establish some parallelisms with previous models  devoted to qubit controlled by an electromagnetic field\cite{Kayanuma2000, Wubs, Child,  Kayanuma2001, Mullen, Sarr2017} and implement new results in the absence of magnetic field. Indeed, the choice we have made to ascribe a cosine shape to all the drives (electric fields) allows us to establish several relevant connections with previous models. If for instance, the transverse drive is switched-off ($\mathcal{A}_{f}=0$), the present study reduces to the one in  Ref.[\onlinecite{Kayanuma2000}]. Our results in Section \ref{Sec3} are in agreement with those of that reference  while the results in Section \ref{Sec4} (Bloch picture investigation) yield additional information. If instead, the longitudinal drive is switched-off ($A=0$), we return to  Ref.[\onlinecite{Wubs}] (for an analytic treatment)  by equally switching off the transverse component ($\Delta=0$) of the magnetic field or to Ref.[\onlinecite{Sarr2017}] (for a numerical treatment) by maintaining it ($\Delta\neq0$).

If now instead of turning off the periodic drives, we rather do so with the magnetic field, depending on the phase shift difference $\phi$, the two situations respectively faced  in [\onlinecite{Child}] and [\onlinecite{Kayanuma2001}]  are met: If we allow the two drives to operate with a phase shift difference  $\phi=0$, we are transported to Ref.[\onlinecite{Child}] and if instead, $\phi=\pi/2$ we move to Ref.[\onlinecite{Kayanuma2001}]. Our analytical results cannot be applied in these cases given that we have in general assumed $v>0$. However, these situations remain interesting and deserve our attention. We wish to quickly implement some relevant results for the cases ($\omega_{f}=\omega$, $\phi=0$) and ($\omega_{f}=2\omega$, $\phi=\pi/2$). Let us recall that the model of interest reads $\mathcal{H}(t)=\frac{A}{2}\cos(\omega t)\boldsymbol{\mathrm{\sigma}}_{z}+\frac{\mathcal{A}_{f}}{2}\cos(\omega_{f} t-\phi)\boldsymbol{\mathrm{\sigma}}_{x}$. We adopt the change of variable $\tau=\sin(\omega t)$. For the first case, this transforms the TDSE to 
 the Rabi problem $i\frac{\partial \ket{\Phi(z)}}{\partial z}=[\frac{A}{2\omega}\boldsymbol{\mathrm{\sigma}}_{z}+\frac{\mathcal{A}_{f}}{2\omega}\boldsymbol{\mathrm{\sigma}}_{x}]\ket{\Phi(z)}$. After solving this equation for the probability amplitude, we obtain
 \begin{eqnarray}\label{e3}
 P_{\uparrow\to\downarrow}(t)=\frac{\mathcal{A}_{f}^{2}}{A^{2}+\mathcal{A}_{f}^{2}}\sin^{2}\Big[\frac{\sqrt{A^{2}+\mathcal{A}_{f}^{2}}}{2\omega}\sin(\omega t)\Big],
 \end{eqnarray} 
 and
 \begin{eqnarray}\label{e4}
 P_{\uparrow\to\uparrow}(t)=\frac{A^{2}}{A^{2}+\mathcal{A}_{f}^{2}}+\frac{\mathcal{A}_{f}^{2}}{A^{2}+\mathcal{A}_{f}^{2}}\cos^{2}\Big[\frac{\sqrt{A^{2}+\mathcal{A}_{f}^{2}}}{2\omega}\sin(\omega t)\Big].
 \end{eqnarray}
 
For the second case, the change of variable leads us to the inverse LZSM problem
$i\frac{\partial \ket{\Phi(\tau)}}{\partial \tau}=\mathcal{H}(\tau)\ket{\Phi(\tau)}$ where 
 $\mathcal{H}(\tau)=\frac{A}{2\omega}\boldsymbol{\mathrm{\sigma}}_{z}+\frac{\mathcal{A}_{f}\tau}{\omega}\boldsymbol{\mathrm{\sigma}}_{x}$. The resulting TDSE is rotated with the help of $\mathbf{R}=\exp(-i\pi\boldsymbol{\mathrm{\sigma}}_{y}/4)$ and we recover the traditional LZSM problem $\mathbf{R}^{\dagger}\mathcal{H}(\tau)\mathbf{R}=-\frac{A}{2\omega}\boldsymbol{\mathrm{\sigma}}_{x}+\frac{\mathcal{A}_{f}\tau}{\omega}\boldsymbol{\mathrm{\sigma}}_{z}$. In the $\tau$-basis, the model has an exact solution. Comparing this with the LZSM model (\ref{equ2}), it is obvious that $v=2\mathcal{A}_{f}/\omega$, $\Delta=-A/\omega$ and the Landau-Zener parameter reads $\delta=\Delta^2/4v$. Starting off at time $t_i=0$ (with $\tau(0)=0$) and ending at a final time $t_f>0$, the propagator of such an evolution is different from (\ref{equ19}) (see Ref.\onlinecite{Torosov} for detailed technique of construction) and yield the measured populations
\begin{eqnarray}\label{e5}
P_{\uparrow\to\downarrow}=\Big({\rm Re} a(t_f,t_i)\Big)^2+\Big({\rm Re} b(t_f,t_i))\Big)^2,
\end{eqnarray} 
and
\begin{eqnarray}\label{e6}
P_{\uparrow\to\uparrow}=\Big({\rm Im} a(t_f,t_i))\Big)^2+\Big({\rm Im} b(t_f,t_i))\Big)^2.
\end{eqnarray}
Here, $a$ and $b$ are Caley-Klein parameters in Eqs.(\ref{equ20}) and (\ref{equ21}) with $z=\sqrt{v}\sin(\omega t) e^{-i\pi/4}$. Eqs.(\ref{e5}) and (\ref{e6}) are yet other contributions to the work in [\onlinecite{Kayanuma2001}] for the case $\omega_{f}=2\omega$ which was not considered there. The results of this section are numerically tested [but not shown] and have depicted a perfect with  agreement numerical data. We conclude that taking separately the models in Ref.[\onlinecite{Kayanuma2000, Wubs, Child, Kayanuma2001, Mullen, Sarr2017}], our model Eq.(\ref{equ1}) is quite rich as compared to each of them.

\section{Conclusion}\label{Sec6}

We have considered a two-level system (TLS) subjected to a linearly polarized magnetic field which substantially changes its sign at time $\tau=0$ in the longitudinal direction where it is equally unbounded and in addition, never turns off in its single transverse direction. The TLS is simultaneously periodically driven in both longitudinal and transverse directions with respectively a RF and a MW signal. A special attention is granted to spin qubit due to versatile potential applications in quantum technology, cryptography, metrology etc. It is on the other hand underlined that the presented results find applications in a wide range of two-level systems from electron spins, atoms, to nuclear (nuclear magnetic resonance) etc. We have considered the ideal situation when the qubit is isolated from unwanted external nuisances. The various control fields applied may be used in realistic situations (where hyperfine and/or spin-orbit interactions prevail) to nearly-decouple the qubit from its environment and minimize the effects of nuclear-spin interactions (spin and boson baths). The coherence time (i.e. the time the qubit preserves information) may be improved. 

The dynamics of the TLS is investigated in two equivalent bases: The Schr\"odinger and Bloch pictures. In the first basis, we evaluate the probabilities of non-adiabatic transitions in the weak and strong driving regimes of the longitudinal drive (RF field). We have established analytical expressions for survival and transition probabilities. Our analytical results are tested by comparing them with data of numerical calculations obtained by directly integrating the time-dependent Schr\"odinger equation. This basis stood out to be appropriate for such a task. In order to gain more insight into the dynamics of the spin qubit under the driving protocol proposed here, we move on to the Bloch picture. Therein, we reduced the optical Bloch vector to three characteristic integral-differential equations: two for coherence factors (spin polarizations) and one for population difference (population inversion). The followings are integrated in the weak driving regime of all the fields. Our analytical results are tested with numerics steaming from the von Neuman equation. Excellent agreements are observed. We have argued that the presented results are adapted for spin manipulations and a few cases are discussed.

\section*{Acknowledgments}

MBK is grateful to AIMS-Ghana for the warm hospitality and where this project started. MBK is thankful to Matteo Smerlak for the wonderful hospitality at Perimeter Institute for Theoretical Physics where part of this project was written. We appreciate useful comments from Sigmund Kohler. We equally thank L. C. Kavi for careful reading of the manuscript and useful languistic suggestions.

\appendix
\section{Additional Properties}\label{App}
We present additional properties of the functions 
\begin{eqnarray}\label{A01}
\mathsf{L}(x,y)=\mathcal{C}(x)\sin(y)-\cos(y)\mathcal{S}(x),
\end{eqnarray}
and
\begin{eqnarray}\label{A02}
\mathsf{M}(x,y)=\mathcal{C}(x)\cos(y)+\sin(y)\mathcal{S}(x),
\end{eqnarray}
namely,
\begin{eqnarray}\label{A1}
\mathsf{L}(x,y)\mathsf{L}(x',y')=\cos(y-y')F_{+}(x,x')-\cos(y+y')F_{-}(x,x')-\sin(y+y')G_{+}(x,x')-\sin(y-y')G_{-}(x,x'),
\end{eqnarray}
and
\begin{eqnarray}\label{A2}
\mathsf{M}(x,y)\mathsf{M}(x',y')=\cos(y-y')F_{+}(y,y')+\cos(y+y')F_{-}(x,x')+\sin(y+y')G_{+}(x,x')-\sin(y-y')G_{-}(x,x'),
\end{eqnarray}
where 
\begin{eqnarray}\label{equ4.6}
F_{\pm}(x,x')=\frac{1}{2}\Big[\mathcal{C}(x)\mathcal{C}(x')\pm \mathcal{S}(x)\mathcal{S}(x')\Big],
\end{eqnarray}
and
\begin{eqnarray}\label{equ4.7}
G_{\pm}(x,x')=\frac{1}{2}\Big[\mathcal{C}(x)\mathcal{S}(x')\pm \mathcal{S}(x)\mathcal{C}(x')\Big].
\end{eqnarray}
Remark
\begin{eqnarray}\label{A3}
\mathsf{L}(x,y)\mathsf{L}(x',y')+\mathsf{M}(x,y)\mathsf{M}(x',y')=2\cos(y-y')F_{+}(x,x')-2\sin(y-y')G_{-}(x,x').
\end{eqnarray}
This property was used to obtain Eq.(\ref{equ4.3}). Similarly,
\begin{eqnarray}\label{A4}
\mathsf{L}(x,y)\mathsf{L}(x',y')-\mathsf{M}(x,y)\mathsf{M}(x',y')=-2\cos(y+y')F_{-}(x,x')-2\sin(y+y')G_{+}(x,x').
\end{eqnarray}
It can also be shown that by shifting the second argument in (\ref{A01}) and (\ref{A02}) this transforms the functions as
\begin{eqnarray}\label{A5}
\mathsf{L}(x,y+y')=\cos(y')\mathsf{L}(x,y)+\mathsf{M}(x,y)\sin(y'),
\end{eqnarray}
and
\begin{eqnarray}\label{A6}
\mathsf{M}(x,y+y')=\cos(y')\mathsf{M}(x,y)-\mathsf{L}(x,y)\sin(y').
\end{eqnarray}
When $x=x'$ and $y=y'$, one can prove from Eqs.(\ref{A3}) and (\ref{A4}) that 
$\mathsf{L}(x,y)=[F_{+}(x,x)-\cos2y F_{-}(x,x)-\sin2y G_{+}(x,x)]^{1/2}$ and 
$\mathsf{M}(x,y)=[F_{+}(x,x)+\cos2y F_{-}(x,x)+\sin2y G_{+}(x,x)]^{1/2}$.
The derivative properties of $\mathsf{L}(x,y)$ or $\mathsf{M}(x,y)$ read
\begin{eqnarray}\label{A7}
\frac{d\mathsf{L}(x,y)}{dy}=\mathsf{M}(x,y), \quad \frac{d\mathsf{M}(x,y)}{dy}=-\mathsf{L}(x,y).
\end{eqnarray}
Therefore, 
\begin{eqnarray}\label{A8}
\frac{d^2\mathsf{L}(x,y)}{dy^2}+\mathsf{L}(x,y)=0, \quad \frac{d^2\mathsf{M}(x,y)}{dy^2}+\mathsf{M}(x,y)=0.
\end{eqnarray}


\section*{References}
\begin{thebibliography}{}
	

\bibitem{lan}
L. D. Landau, Phys. Z. Sowietunion {\bf 2},  46 (1932).

\bibitem{zen}
C. Zener, Proc. R. Soc. A. {\bf 137}1, 696 (1932).

\bibitem{stu}
E. C. G. St\"uckelberg, Helv. Phys.  Acta  {\bf 5}, 369 (1932).

\bibitem{Majorana}
E. Majorana, Nuovo Cimento {\bf 9}, 43 (1932).


\bibitem{Bulka}
J. Luczak, B. R. Bulka, Quantum Inf Process (2017) 16:10.


\bibitem{Cao}
G. Cao, H.-O. Li, T. Tu, L. Wang, C. Zhou, M. Xiao, G.-C. Guo, H.-W. Jiang, and G.-P. Guo, Nature Communications {\bf 4}, 1401 (2013).

\bibitem{Ribeiro}
H. Ribeiro, J. R. Petta, and G. Burkard, Phys. Rev. B {\bf 82}, 115445 (2010).

\bibitem{Petta}
J. R. Petta, H. Lu, and A. C. Gossard, Science {\bf 327},  669 (2010).

\bibitem{Shev}
S. N. Shevchenko, S. Ashhab, and F. Nori, Physics Reports {\bf 492}, 1 (2010).

\bibitem{Oliver}
W. D. Oliver, Y. Yu, J. C. Lee, K. K. Berggren, L. S. Levitov, and T. P. Orlando, Science {\bf 310}, 1653 (2005).

\bibitem{Saito2007}
K. Saito, M. Wubs, Sigmund Kolher, Y. Kayanuma, P. Hanggi, Phys. Rev B {\bf 75}, 214308 (2007).

\bibitem{Wernsd}
W. Wernsdorfer, R. Sessoli, A. Caneschi, D. Gatteschi and A. Cornia, Euro. Phys. Lett. {\bf 50}, 552 (2000).

\bibitem{Sillanp}
M. Sillanp\"a\"a, T. Lehtinen, A. Paila, Y. Makhlin, and P. Hakonen, Phys. Rev. Lett. {\bf 96}, 187002 (2006).


\bibitem{Petta2012}
J. Stehlik, Y. Dovzhenko, J. R. Petta, J. R. Johansson, F. Nori, H. Lu, and A. C. Gossard, Phys. Rev. B {\bf 86}, 121303(R) (2012).

\bibitem{sun2016}
M. Gong, Y. Zhou, D. Lan, Y. Fan, J. Pan,   H. Yu, J. Chen, G. Sun, Y. Yu,   S. Han, and P. Wu, 
Appl. Phys. Lett. {\bf 108}, 112602 (2016).



\bibitem{Vitanov1997}
B. M. Garraway and N. V. Vitanov, Phys. Rev. A {\bf 55}, 4418 (1997).

\bibitem{Kayanuma2000}
Y. Kayanuma and Y. Mizumoto, Phys. Rev. A {\bf 62} , 061401(R) (2000);
D. Suqing, L.-B. Fu, J. Liu, X.-G. Zhao, Physics Letters A {\bf 346}, 315-320 (2005).

\bibitem{Wubs}
M. Wubs, K. Saito, S. Kohler, Y. Kayanuma and P. H\"anggi, New Journal of Physics {\bf7} 218 (2005).

\bibitem{Child}
L. Childress and J. McIntyre, Phys. Rev. A {\bf82}, 033839 (2010).

\bibitem{Kayanuma2001}
K.-I. Noba and Y. Kayanuma, Phys. Rev. A {\bf64}, 013413 (2001).

\bibitem{Grossmann}
F. Grossmann, T. Dittrich, P. Jung, and P. H\"anggi, Phys. Rev. Lett. {\bf 67}, 516 (1991).

\bibitem{Mullen}
K. Mullen, E. Ben-Jacob, Y. Gefen, and Z. Schuss, Phys. Rev. Lett.
{\bf62}, 2543 (1989).

\bibitem{Sarr2017}
F. Sarreshtedari and M. Hosseini, Phys. Rev. A {\bf95}, 033834 (2017).

\bibitem{Viatnov1996}
N. V. Vitanov, B. M. Garraway, Phys. Rev. A {\bf53}, 4288 (1996).

\bibitem{KenmoePRB2}
M. B. Kenmoe, A. B. Tchapda, and L. C. Fai, arXiv:1703.06970v2 (2017).

\bibitem{Book}
A. Erdelyi, W. Magnus, F. Oberhettinger, and F. G. Tricomi, Higher Transcendental Functions, Vol.{\bf 2} (McGraw-Hill, New
York, 1953).

\bibitem{Cohen}
C. C.-Tannoudji, B. Diu, F. Laloe, Quantum Mechanics, Vol.{\bf 1} (Wiley-Science, 1977).

\bibitem{Guido}
H. Ribeiro, J. R. Petta, G. Burkard, Phys. Rev. B {\bf 82}, 115445 (2010).


\bibitem{pok2007}
V. L. Pokrovsky and D. Sun, Phys. Rev. B 76,  024310 (2007).

\bibitem{Kenmoe2015}
M. B. Kenmoe, S. E. Mkam Tchouobiap, J. E. Danga, C. Kenfack Sadem, L. C. Fai, Physics Letters A {\bf 379}, 635-642 (2015).

\bibitem{Muga}
A. Ruschhaupt, X. Chen, D. Alonso, and J. G. Muga, New Journal of Physics {\bf 14}, 093040 (2012).

\bibitem{Sinitsyn}
C. Sun, A. Saxena, and N. A. Sinitsyn, arXiv:1703.10271v1 (2017).

\bibitem{Torosov}
B. T. Torosov, and N. V. Vitanov, J. Phys. A: Math. Theor. {\bf 41}, 155309 (2008).

\end{thebibliography}
\end{document}